\documentclass[a4paper,fleqn,usenatbib]{mnras}

\usepackage{newtxtext,newtxmath}
\usepackage[T1]{fontenc}

\usepackage[english]{babel}
\usepackage{graphicx}
\usepackage{fleqn}
\usepackage{amssymb}
\usepackage{amsmath}
\usepackage{booktabs}
\usepackage{hyperref}
\usepackage{comment}

\usepackage{color, colortbl}

\newcommand*{\mycdot}{\kern-.2em\cdot\kern-.2em}
\renewcommand{\S}{Section}
\newcommand{\F}{Fig.}
\newcommand{\codename}{SecularMultiple}

\newcommand{\msun}{\mathrm{M}_\odot}
\newcommand{\ud}{\mathrm{d}}
\newcommand{\uyr}{\mathrm{yr}}
\newcommand{\rsun}{\mathrm{R}_\odot}

\newcommand{\au}{\,\textsc{au}}
\newcommand{\renc}{R_\mathrm{enc}}

\pubyear{2018}

\allowdisplaybreaks

\title[Shrinking quadruples]{Shrinking orbits in hierarchical quadruple star systems}

\author[Hamers]{Adrian S. Hamers$^{1}$\thanks{E-mail: hamers@ias.edu} \\
$^{1}$Institute for Advanced Study, School of Natural Sciences, Einstein Drive, Princeton, NJ 08540, USA}

\date{Accepted 2018 October 20. Received 2018 October 13; in original form 2018 August 13}

\begin{document}

\label{firstpage}
\pagerange{\pageref{firstpage}--\pageref{lastpage}}
\maketitle

\begin{abstract} 
The distribution of the inner orbital periods of solar-type main-sequence (MS) triple star systems is known to be peaked at a few days, and this has been attributed to tidal evolution combined with eccentricity excitation due to Lidov-Kozai oscillations. Solar-type MS quadruple star systems also show peaks in their inner orbital period distributions at a few days. Here, we investigate the natural question whether tidal evolution combined with secular evolution can explain the observed inner orbital period distributions in quadruple stars. We carry out population synthesis simulations of solar-type MS quadruple star systems in both the 2+2 (two binaries orbiting each other's barycentre) and 3+1 (triple orbited by a fourth star) configurations. We take into account secular gravitational and tidal evolution, and the effects of passing stars. We assume that no short-period systems are formed initially, and truncate the initial orbital period distributions below 10 d accordingly. We find that, due to secular and tidal evolution, the inner orbital period distributions develop tails at short periods. Although qualitatively consistent with the observations, we find that our simulated orbital period distributions only quantitatively agree with the observations for the 3+1 systems. The observed 2+2 systems, on the other hand, show an enhancement of systems around 10 d, which is not reproduced in the simulations. This suggests that the inner orbital periods of 2+2 systems are not predominantly shaped by tidal and secular evolution, but by other processes, most likely occurring during the stellar formation and early evolution.
\end{abstract}


\section{Introduction}
\label{sect:introduction}
It has been known for over a decade that the orbital period distribution of isolated solar-type main-sequence (MS) binaries is different from their counterparts with tertiary companions. Specifically, the presence of a tertiary star implies a peak in the period distribution at $\sim 3\,\ud$, which is not present in the distribution of isolated binaries \citep{2002A&A...382..118T,2006A&A...450..681T}. This has been attributed to shrinkage of the binary orbit due to tides enhanced by high eccentricities, which are driven by secular Lidov-Kozai (LK) oscillations (\citealt{1962P&SS....9..719L,1962AJ.....67..591K}; see \citealt{2016ARA&A..54..441N} for a review). This process, known as LK cycles with tidal friction, has been studied in detail by numerous authors \citep{1979A&A....77..145M,2001ApJ...562.1012E,2006Ap&SS.304...75E,2007ApJ...669.1298F,2013MNRAS.430.2262H,2014ApJ...793..137N,2016ComAC...3....6T,2017MNRAS.467.3066A,2018MNRAS.479.4749B}, and may be responsible for producing a large fraction of short-period binaries that would otherwise not be expected to form due to the larger sizes of the stars during the pre-MS.  

Triple systems constitute about 10\% of all stellar systems with solar-type MS stars in the solar neighbourhood (e.g., \citealt{2014AJ....147...86T,2014AJ....147...87T}). A smaller, but non-negligible fraction of stellar systems, about 1\%, is composed of quadruple star systems. The latter are known to occur in the 2+2 (two binaries orbiting each other's barycentre) and 3+1 (triple orbited by a fourth star) configurations. The long-term dynamics of hierarchical quadruple systems are complicated, and have been considered by a number of authors \citep{2013MNRAS.435..943P,2015MNRAS.449.4221H,2016MNRAS.461.3964V,2017MNRAS.470.1657H,2018MNRAS.476.4234F,2018MNRAS.474.3547G,2018MNRAS.478..620H}. In particular, it has been shown that the efficiency to attain high eccentricities in these quadruple systems is higher compared to equivalent triples, i.e., if two stars would be replaced by a single star. 

Similarly to triple stars, the inner orbital period distributions of solar-type MS quadruple stars show an enhancement at a few to several tens of days \citep{2008MNRAS.389..925T}. Therefore, a natural question is whether tidal friction combined with secular evolution can explain this enhancement, in analogy to triple star systems. Naively, one might expect this process to be very efficient given the higher efficiency to attain high eccentricities in quadruple compared to triple stars. However, one should also take into account that the secular evolution of these systems, in the parameter space in which the enhancement is large, is typically chaotic \citep{2015MNRAS.449.4221H,2017MNRAS.470.1657H,2018MNRAS.474.3547G}, and the time-scale for reaching high eccentricities can be long, i.e., longer than the MS lifetimes. 

The aim of this paper is to study the formation of close binaries in solar-type MS quadruple systems through tidal and secular evolution. We assume that no short-period systems are formed initially, and accordingly truncate the initial orbital period distributions below 10 d. We carry out population synthesis simulations of quadruple stars in both the 2+2 and 3+1 configurations, taking into account secular gravitational and tidal evolution, and the effects of passing stars in the field. 

The plan of the paper is as follows. In \S\,\ref{sect:meth}, we describe the numerical algorithm used for our simulations. We illustrate two notable consequences of tidal migration in quadruple systems in \S\,\ref{sect:examples}. The initial conditions and assumptions for the population synthesis simulations are discussed in \S\,\ref{sect:IC}. We present our results in \S\,\ref{sect:pop_syn} and discuss them in \S\,\ref{sect:discussion}, and we conclude in \S\,\ref{sect:conclusions}.

\section{Numerical algorithm}
\label{sect:meth}

\begin{figure}
\center
\includegraphics[scale = 0.65, trim = 0mm 0mm 0mm 0mm]{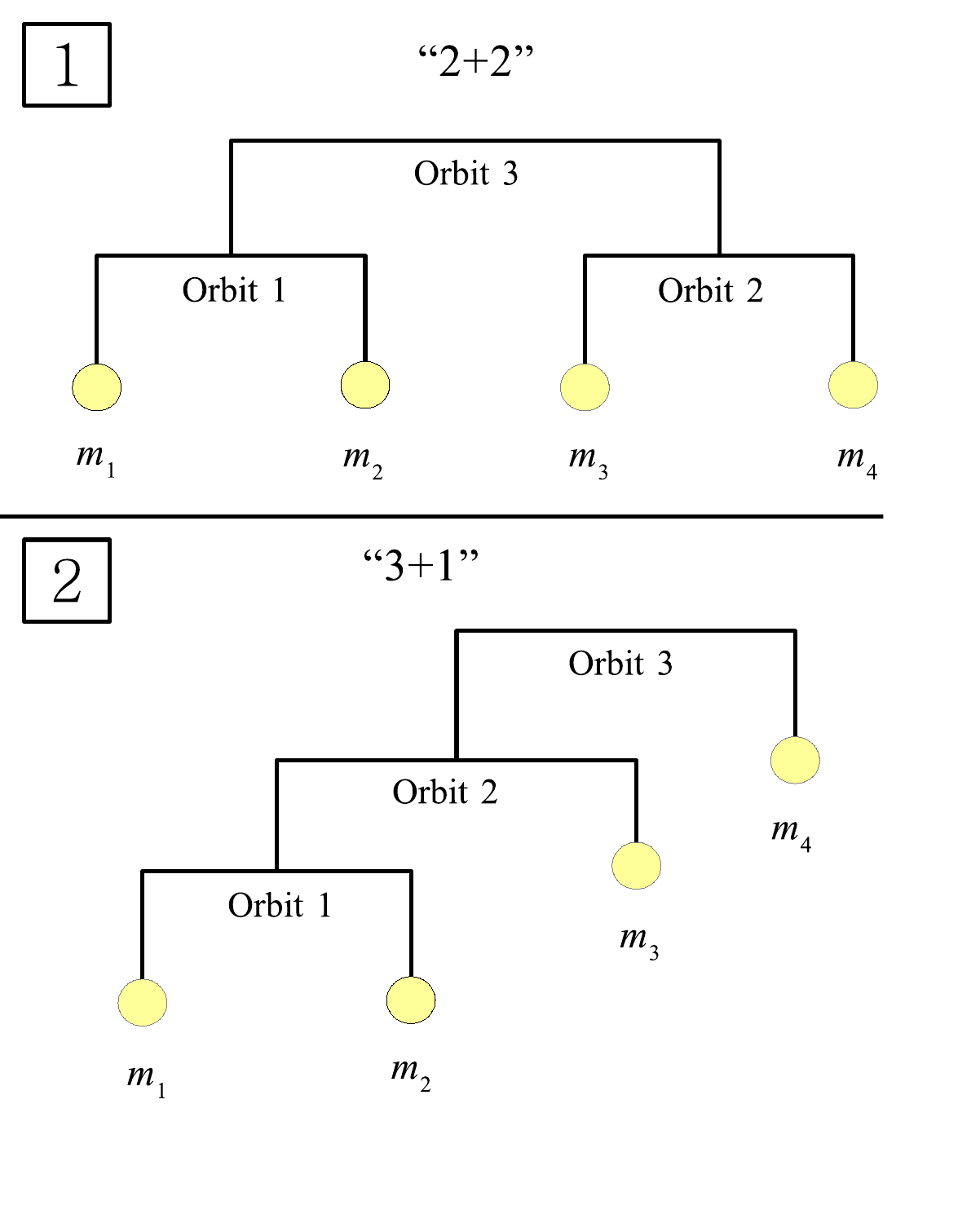}
\caption {Illustration of the types of systems considered in this paper in a mobile diagram \citep{1968QJRAS...9..388E}. Top: the 2+2 configuration; bottom: the 3+1 configuration. }
\label{fig:configurations}
\end{figure}

\begin{table*}
\begin{tabular}{lp{3.5cm}p{12.0cm}}
\toprule
Symbol & Description & Initial value(s) and/or distribution in population synthesis \\
\midrule
{\it Quadruple} \\
{\it  system} \\
$m_1$					& Mass of the primary star in orbit 1.																& $1\,\msun$ \\
$m_2$					& Mass of the secondary star in orbit 1. 															& $m_1 q_1$, where $q_1$ has a flat distribution and with$0.1<m_2/\msun<1$. \\
$m_3$ 					& Mass of star 3.																			& $q_2 (m_1+m_2)$, where $q_2$ has a flat distribution, and with $0.1<m_3/\msun<1$. \\
$m_4$					& Mass of star 4.																			& $q_2 m_3$, and with $0.1<m_4/\msun<1$. \\
$R_i$					& Radius of star $i$.																			& $(m_i/\msun)^{0.8}\,\rsun$ \citep{1994sse..book.....K}\\
$P_{\mathrm{s},i}$			& Spin period of star $i$. 																		& $10\,\ud$ \citep{2007ApJ...669.1298F} \\
$\theta_{\mathrm{s},i}$		& Obliquity (spin-orbit angle) of star $i$. 															& $0^\circ$ \\
$t_{\mathrm{V},i}$			& Viscous time-scale of star $i$. 																& $5\,\uyr$ \citep{2007ApJ...669.1298F} \\
$k_{\mathrm{AM},i}$			& Apsidal motion constant of star $i$.															& 0.014 \citep{2007ApJ...669.1298F} \\
$r_{\mathrm{g},i}$			& Gyration radius of star $i$.																	& 0.08 \citep{2007ApJ...669.1298F} \\
$P_{\mathrm{orb},i}$			& Orbital period of orbit $i$. 																	& (1) Gaussian distribution in $\mathrm{log}_{10}(P_{\mathrm{orb},i}/\uyr)$ with mean 5.03 and standard deviation 2.28 \citep{2010ApJS..190....1R} or (2) an \"{O}pic law (\citealt{opic1924}; flat distribution in $\mathrm{log}_{10} a_i$). In both cases, the orbital periods range between $10$ and $10^9$ d, and the systems are subject to dynamical stability constraints \citep{2001MNRAS.321..398M}, and $a_i(1-e_i^2) > a_\mathrm{crit}$ for orbits 1 and 2 (2+2), or orbit 1 (3+1). Here, $a_\mathrm{crit}$ is the semimajor axis corresponding to an orbital period of 10 d. \\
$a_i$					& Semimajor axis of orbit $i$. 																	& Computed from $P_{\mathrm{orb},i}$ and the $m_i$ using Kepler's law. \\
$e_i$					& Eccentricity of orbit $i$. 																		& (1) Rayleigh distribution between 0.01 and 0.95 with an rms width of 0.33 \citep{2010ApJS..190....1R}. (2) Flat distribution between 0.01 and 0.95. \\
$i_i$						& Inclination of orbit $i$. 																		& $0$-$180^\circ$ (flat distribution in $\cos i_i $) \\
$i_{ij}$					& Inclination of orbit $i$ relative to orbit $j$.														& $0$-$180^\circ$ (flat distribution in $\cos i_{ij} $) \\
$\omega_i$				& Argument of periapsis of orbit $i$. 																& $-180$-$180^\circ$ (flat distribution in $\omega_i$) \\
$\Omega_i$				& Longitude of the ascending node of orbit $i$. 														& $-180$-$180^\circ$ (flat distribution in $\Omega_i$) \\
\midrule
{\it Flybys} \\
$M_\mathrm{per}$			& Mass of the perturbers. 																		& $0.1$-$80\,\msun$ with a Kroupa initial mass function \citep{1993MNRAS.262..545K}, corrected for gravitational focusing and a stellar age of 10 Gyr. \\
$Z_i$					& Metallicity of perturbers.																		& $0.02$ \\
$n_\star$					& Stellar number density. 																		& $0.1 \, \mathrm{pc^{-3}}$ \citep{2000MNRAS.313..209H} \\
$\renc$					& Encounter sphere radius.																	& $0, 10^4\,\au$ \\
$\sigma_\star$				& One-dimensional stellar velocity dispersion.														& $30\,\mathrm{km\,s^{-1}}$ \\
\bottomrule
\end{tabular}
\caption{Description of important quantities and their initial value(s) and/or distributions assumed in the population synthesis (\S\,\ref{sect:pop_syn}). }
\label{table:IC}
\end{table*}

We model the long-term evolution of stellar hierarchical quadruple systems in both the `2+2' (two binaries orbiting each other's barycentre) and `3+1' (a triple orbited by a fourth star) configurations. An illustration of the configurations is given in \F\,\ref{fig:configurations}. We give an overview of our notation in Table~\ref{table:IC}, in which we also summarize in the third column the assumptions made in the population synthesis study (\S s\,\ref{sect:IC} and \ref{sect:pop_syn}). Throughout, we will refer to the `inner' orbits as orbit 1 or 2 for the 2+2 configuration, and orbit 1 for the 3+1 configuration. The `outermost' or `outer' orbit is orbit 3 in both configurations, and the `intermediate' orbit is orbit 2 in the 3+1 configuration. 

Our numerical algorithm is similar to that of \citet{2018MNRAS.478..620H}, with the exception that we do not use a dedicated code to model the stellar evolution. Since our focus is on solar-type MS stars, the stellar parameters (mass and radius) do not change significantly during 10 Gyr, so it is reasonable to assume these parameters to be constant. Below we give a brief description of the numerical algorithm. For more details, we refer to \citet{2018MNRAS.478..620H} and references therein.

\subsection{Secular dynamical evolution}
\label{sect:meth:sec}
To model the secular dynamics, we use \textsc{\codename} \citep{2016MNRAS.459.2827H}, which is a generalization of a code developed earlier for 3+1 quadruple systems \citep{2015MNRAS.449.4221H}. The \textsc{\codename} code is based on an expansion of the Hamiltonian of the system in terms of ratios of separations of orbits on different levels (e.g., in \F\,\ref{fig:configurations}, the small expansion parameters are $r_1/r_3$ and $r_2/r_3$ for the 2+2 configuration, and $r_1/r_2$ and $r_2/r_3$ for the 3+1 configuration, respectively, which $r_i$ is the separation of orbit $i$). The Hamiltonian is subsequently orbit averaged, and the orbit-averaged equations of motion are solved numerically. In the integrations, we include terms up to and including octupole order (third order in the separation ratios) for interactions involving three binaries, and up to and including dotriacontupole order (fifth order in the separation ratios) for pairwise interactions. 

Post-Newtonian (PN) corrections are included in each orbit to the 1 and 2.5PN orders (i.e., including relativistic precession, and energy and angular-momentum loss due to gravitational wave radiation). Any `cross' terms, i.e., PN terms involving more than one orbit simultaneously \citep{2013ApJ...773..187N}, are neglected. We note that although they are included for completeness, the 2.5PN terms are not important for solar-type MS stars. The 1PN terms can be important by quenching secular evolution due to apsidal precession in the inner orbits. 

\subsection{Tidal evolution}
\label{sect:meth:tides}
Tidal evolution is modelled with the equilibrium tide model \citep{1981A&A....99..126H,1998ApJ...499..853E}. Specifically, we use equations (81) and (82) of \citet{1998ApJ...499..853E}, with the non-dissipative terms $X$, $Y$ and $Z$ given explicitly by equation (10)-(12) of \citet{2001ApJ...562.1012E}, and the dissipative terms $V$ given explicitly by equations (A7)-(A11) of \citet{2009MNRAS.395.2268B}. In these equations, we include the effects of dissipative tides, orbital precession due to tidal bulges, and orbital precession due to stellar rotation (assuming uniform rotation, but allowing the spins to be non-parallel to the orbit). The spin vectors of all stars are tracked and the spins are not confined to be parallel with the orbit, although we initialize the spins to be parallel with the orbit (i.e., the initial obliquity $\theta_{\mathrm{s},i}=0^\circ$), with a spin period of $P_{\mathrm{s},i} = 10\,\ud$. 

We assume a constant viscous time-scale $t_{\mathrm{V},i}$ for the stars, which we set to $t_{\mathrm{V},i}=5\,\uyr$. The apsidal motion constants are set to $k_{\mathrm{AM},i}=0.014$, and the gyration radii are set to $r_{\mathrm{g},i}=0.08$ (i.e., the moment of inertia is $I_i = r_{\mathrm{g},i} m_i R_i^2$). The values for $P_{\mathrm{s},i}$, $t_{\mathrm{V},i}$, $k_{\mathrm{AM},i}$, and $r_{\mathrm{g},i}$ are adopted from \citet{2007ApJ...669.1298F}.

We remark that many uncertainties exist regarding the efficiency of tidal dissipation (see, e.g., \citealt{2014ARA&A..52..171O} for a review). In particular, we do not take into account dynamical tides, which could be important at high eccentricities, when the orbits are close to parabolic (\citealt{1977ApJ...213..183P}; the transition eccentricity is around 0.8, \citealt{1995ApJ...450..732M}). A more sophisticated tidal model, e.g., in which the equilibrium tide is used for low eccentricities and the formalism of \citet{1977ApJ...213..183P} for high eccentricities such as in \citet{2018ApJ...854...44M}, is beyond the scope of this paper.

\subsection{Flybys}
\label{sect:meth:flybys}
We include the gravitational effects of stars passing by the quadruple system using the impulsive approximation, i.e., the stars in the quadruple system can be considered to be fixed in space whereas the perturber imparts velocity kicks on each of the components. These kicks lead to changes of the orbits, in principle affecting all orbital elements, but most significantly the semimajor axes and eccentricities of wide orbits ($a_i\gtrsim 10^4\,\au$). Our method for computing the effects of impulsive flybys on the orbits of the system is the same as in \citet{2017AJ....154..272H} and \citet{2018MNRAS.478..620H}; for details, we refer to the latter papers. 

We adopt the same parameters as in \citet{2018MNRAS.478..620H}, i.e., we assume a locally homogeneous stellar background with a number density $n_\star = 0.1 \, \mathrm{pc^{-3}}$ \citep{2000MNRAS.313..209H}, a one-dimensional velocity dispersion $\sigma_\star = 30\,\mathrm{km\,s^{-1}}$, and a Kroupa mass function \citep{1993MNRAS.262..545K} between 0.1 and 80 $\msun$, corrected for gravitational focusing and stellar evolution. The correction for stellar evolution is carried out by replacing the initial mass with the final mass after 10 Gyr of stellar evolution assuming solar metallicity and using the \textsc{SeBa} stellar evolution code \citep{1996A&A...309..179P,2012A&A...546A..70T} in \textsc{AMUSE} \citep{2013CoPhC.183..456P,2013A&A...557A..84P}. The radius of the sphere used to compute the encounter rate and encounter properties, $\renc$, is set to either $\renc=0$ (no encounters), or $\renc = 10^4\,\au$, such that most encounters with the widest orbit (orbit 3) are impulsive, whereas not too large as to be computationally too inhibitive. We reject sampled encounters that are secular in nature (i.e., the speed of the perturber is much slower than the orbital speed in the quadruple system), since the effects of secular encounters are typically negligible compared to those of impulsive encounters (Hamers, unpublished). We neglect the effects of encounters in the intermediate regime between secular and impulsive encounters, although some encounters in this regime could be important \citep{2010ApJ...725..353D}. 

\subsection{Stopping conditions}
\label{sect:meth:sc}
We impose a number of stopping conditions in our simulations. 
When high eccentricities are reached in the inner orbits, Roche lobe overflow (RLOF) can be triggered, leading to mass transfer, and possibly to the coalescence of two stars. If the mass transfer and secular time-scales are comparable, then the resulting evolution can be complicated, since the two effects are competing (RLOF tends to circularize the orbits, and secular evolution tends to increase the eccentricity). This complicated process is beyond the scope of the paper. Instead, we simply check for the onset of RLOF in eccentric orbits, and stop the simulation. 

Furthermore, we stop the simulation when the system becomes dynamically unstable. This is most likely to occur in 3+1 systems, when orbit 2 is driven to high eccentricity due to the torque of orbit 3, driving dynamical instability of the orbit pair 1-2. If a dynamical instability is triggered, then stars may collide or could become ejected from the system, destroying the hierarchy of the system. 

The implemented stopping conditions are as follows.
\begin{enumerate}
\item One of the stars fills its Roche lobe, either after tidal evolution in a circular orbit, or in an eccentric orbit. We use the fits of \citet{2007ApJ...660.1624S}, in particular, equations~(47) through (52) evaluated at periapsis, to determine the instantaneous Roche lobe radius. The latter equations give the Roche lobe radius as a function of orbital phase, spin frequency, eccentricity and mass ratio, as a correction to the Roche lobe radius fits of \citet{1983ApJ...268..368E} evaluated at periapsis (i.e., $a$ in \citealt{1983ApJ...268..368E} is replaced by $a[1-e]$). We note that, for MS stars, RLOF always occurs prior to a physical collision.
\item One of the orbit pairs becomes dynamically unstable. To evaluate stability, we use the criterion of \citet{2001MNRAS.321..398M}, which is assumed to be correct for quadruple systems if two of the bodies are appropriately replaced by a single body. In the case of 2+2 systems, we apply the stability criterion to the 1-3 and 2-3 orbit pairs; in the case of 3+1 systems, we apply the stability criterion to the 1-2 and 2-3 orbit pairs (see also \S\,\ref{sect:IC:orbits}). We remark that a breakdown of the secular equations of motion could already occur before dynamical stability. This caveat is discussed in \S\,\ref{sect:discussion:sub}.
\item The age of the system exceeds 10 Gyr.
\item The CPU wall time exceeds 24 hr. Some systems, in particular for the 3+1 configuration, take excessively long to integrate due to very short secular time-scales compared to 10 Gyr. Although we terminate the integration of these systems for practical reasons, we show below in \S\,\ref{sect:discussion:exceed} that the majority of these systems are not expected to lead to tidal migration, and the stopping condition therefore does not strongly affect our results.
\end{enumerate}

We note that conditions (i) and (ii), which depend on the instantaneous orbital semimajor axes and eccentricities, are implemented as root finding conditions within the set of ordinary differential equations (which are solved using the \textsc{CVODE} routine, \citealt{1996ComPh..10..138C}, which supports root finding). Therefore, there is no risk of the stopping conditions being missed in the simulations due to a finite number of output snapshots.

\section{Examples of tidal migration}
\label{sect:examples}
Here, we discuss two notable examples of tidal migration as found in the population synthesis simulations (\S\,\ref{sect:pop_syn}). The initial conditions of the two examples are listed in Table\,\ref{table:IC_example}. For other parameters such as the viscous time-scale, we refer to Table\,\ref{table:IC}.

\begin{table*}
\begin{tabular}{ccccccccccccccccccc}
\toprule
$m_1$ & $m_2$ & $m_3$ & $m_4$ & $a_1$ & $a_2$ & $a_3$ & $e_1$ & $e_2$ & $e_3$ & $i_1$ & $i_2$ & $i_3$ & $\omega_1$ & $\omega_2$ & $\omega_3$ & $\Omega_1$ & $\Omega_2$ & $\Omega_3$ \\
$\msun$ & $\msun$ & $\msun$ & $\msun$ & $\au$ & $\au$ & $\au$ & & & &deg & deg & deg & deg & deg & deg & deg & deg & deg \\
\midrule
1.0 & 0.7 & 0.4 & 0.1 & 3.6 & 0.1 & 21.3 &0.08 & 0.23 & 0.32 & 13.8 & 55.5 & 102.0 & 59.2 & -77.9 & -94.6 & 169.6 & 6.5 & -128.6 \\
\midrule
1.0 & 0.2 & 0.9 & 0.7 & 0.7 & 82.8 & 5949.4 & 0.50 & 0.42 & 0.24 & 129.6 & 67.7 & 115.5 & 82.0 & 149.5 & -106.8 & 40.5 & 85.9 & 153.7 \\
\bottomrule
\end{tabular}
\caption{ Initial conditions for the two example systems discussed in \S\,\ref{sect:examples}. The first (second) row of entries corresponds to the example shown in \F\,\ref{fig:example1} (\ref{fig:example2}). }
\label{table:IC_example}
\end{table*}

\begin{figure}
\center
\includegraphics[scale=0.47,trim = 5mm 10mm 0mm 0mm]{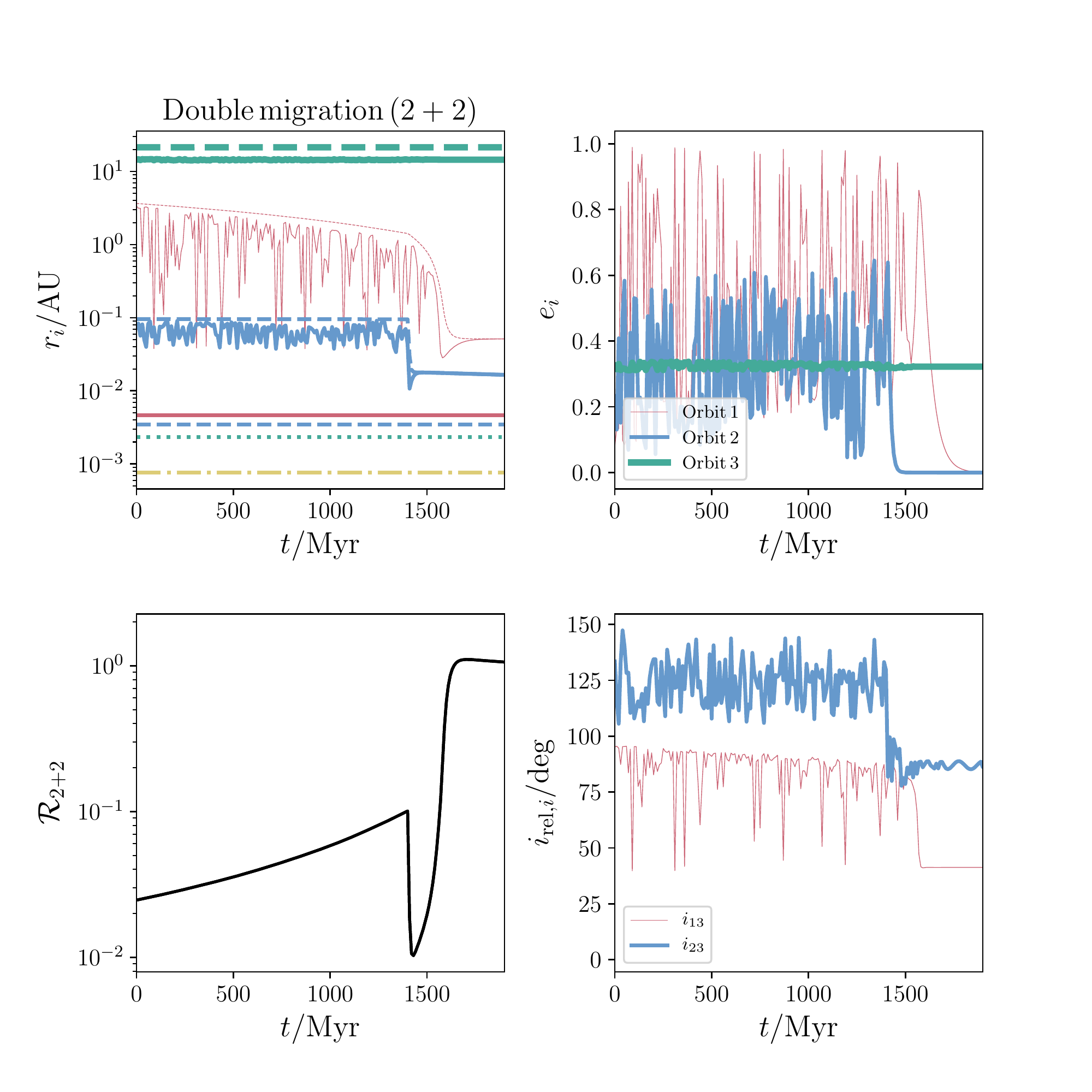}
\caption { Example evolution of a 2+2 system in which both inner orbits migrate to short periods. The initial conditions are given in the first row of entries in Table~\ref{table:IC_example}. Top-left panel: the semimajor axes (dashed lines), periapsis distances $a_i(1-e_i)$ (solid lines), and stellar radii (solid, dashed, dotted, and dot-dashed lines for stars 1 through 4) as a function of time. For the orbits, thin red, thick blue and very thick green lines correspond to orbits 1, 2, and 3, respectively; for the stars, red, blue, green, and yellow lines correspond to stars 1, 2, 3 and 4, respectively. Top-right panel: eccentricities of the orbits, using the same colours and thicknesses as in the top-left panel. Bottom-left panel: the ratio of LK time-scales equation~(\ref{eq:R_2p2}). Bottom-right panel: the relative inclinations of orbits 1 and 2 to their parent (i.e., $i_{13}$, thin red line, and $i_{23}$, thick blue line). Refer to \S\,\ref{sect:examples:1} for a detailed description of the evolution. }
\label{fig:example1}
\end{figure}

\subsection{Double migration in a 2+2 system}
\label{sect:examples:1}
First, we show in \F\,\ref{fig:example1} an example for the 2+2 configuration in which `double migration' occurs, i.e., both orbits 1 and 2 shrink to $P_{\mathrm{orb,i}}<10\,\ud$. In the figure, we plot, as a function of time in the top-left panel, the semimajor axes (dashed lines), periapsis distances $a_i(1-e_i)$ (solid lines), and stellar radii (solid, dashed, dotted, and dot-dashed lines for stars 1 through 4). The top-right panel shows the eccentricities, the bottom-left panel shows the ratio of LK time-scales, i.e., \citep{2017MNRAS.470.1657H}
\begin{align}
\label{eq:R_2p2}
\mathcal{R}_{2+2} &\equiv \frac{t_\mathrm{LK,13}}{t_\mathrm{LK,23}} \simeq \left ( \frac{a_2}{a_1} \right)^{3/2} \left ( \frac{m_1+m_2}{m_3+m_4} \right )^{3/2},
\end{align}
and the bottom-right panel shows the mutual inclinations of orbit pairs 1-3 and 2-3. We note that, when $\mathcal{R}_{2+2}$ is close to unity (roughly speaking, within an order of magnitude), secularly chaotic behaviour and particularly high eccentricities are to be expected (\citealt{2017MNRAS.470.1657H}; similar behaviour applies to the 3+1 configuration in terms of $\mathcal{R}_{3+1}$, see the second example below in \S\,\ref{sect:examples:2}).

Initially, orbit 1 is highly inclined with respect to orbit 3, with $i_{13,\mathrm{i}} \simeq 95.2^\circ$. This triggers high-amplitude eccentricity oscillations in orbit 1, which cause its semimajor axis to gradually decrease. Consequently, the ratio $\mathcal{R}_{2+2}$ gradually increases. Initially, orbit 2 is inclined with respect to orbit 3 by $i_{23,\mathrm{i}} \simeq 133.4^\circ$, and the resulting maximum eccentricities in orbit 2 are not sufficiently high to trigger tidal migration. However, at $\simeq 1400\,\mathrm{Myr}$, due to the tidal migration of orbit 1, $\mathcal{R}_{2+2}$ reaches $\sim 0.1$, and the two orbits (1 and 2) become dynamically coupled due to secular resonances. Consequently, the eccentricity is excited in orbit 2, causing tidal migration in that orbit, and which initially decreases $\mathcal{R}_{2+2}$. As orbit 1 continues to shrink, $\mathcal{R}_{2+2}$ increases again. Finally, the orbits circularize due to tides with $a_{1,\mathrm{f}} \simeq 4\times 10^{-2}\,\au$, and $a_{2,\mathrm{f}} \simeq 1\times 10^{-2}\,\au$. This example shows that tidal shrinkage of one orbit can trigger the second orbit to shrink as well. Due to the gradual shrinkage of the first orbit, secular resonances are almost guaranteed to occur. Note, however, that the occurrence of this phenomenon does require $\mathcal{R}_{2+2}$ to pass within roughly an order of magnitude during tidal migration.

\begin{figure}
\center
\includegraphics[scale = 0.47, trim = 5mm 10mm 0mm 0mm]{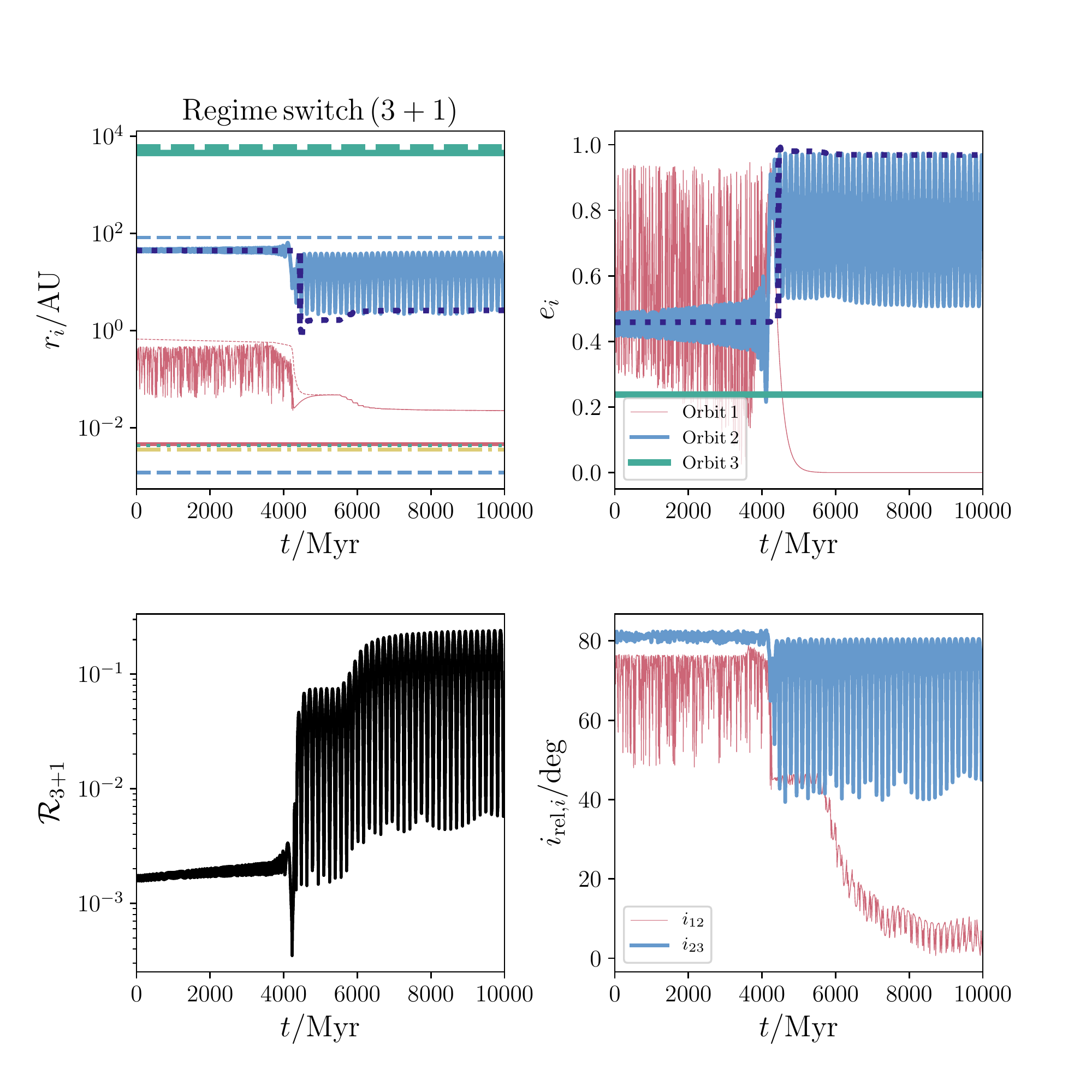}
\caption {Example of a 3+1 system in tidal migration of the innermost orbit brings the system into a different dynamical regime, resulting in enhanced eccentricities of orbit 2. The initial conditions are given in the second row of entries in Table~\ref{table:IC_example}. See the caption of \F\,\ref{fig:example1} for an explanation of the colours and line styles used, and \S\,\ref{sect:examples:2} for a detailed description of the evolution. Here, the bottom-right panel shows the ratio of LK time-scales equation~(\ref{eq:R_3p1}). In the top-left and top-right panels, the dark blue-dotted lines show the periapsis distance and eccentricity of orbit 2, respectively, based on a semianalytic calculation of the maximum eccentricity of orbit 2 using the method described in section 3.4.2 of \citet{2015MNRAS.449.4221H}. }
\label{fig:example2}
\end{figure}

\subsection{Switching of dynamical regimes due to migration in a 3+1 system}
\label{sect:examples:2}
In the second example, we show how tidal migration in the inner orbit of a 3+1 system can bring the system into a different dynamical regime. We show the evolution in \F\,\ref{fig:example2}, which is similar to \F\,\ref{fig:example1}, except that the bottom-left panel now shows the ratio of LK time-scales \citep{2017MNRAS.470.1657H}
\begin{align}
\label{eq:R_3p1}
\mathcal{R}_{3+1} &\equiv \frac{t_\mathrm{LK,12}}{t_\mathrm{LK,23}} \simeq \left ( \frac{a_2^3}{a_1 a_3^2} \right)^{3/2} \left ( \frac{m_1+m_2}{m_1+m_2 + m_3} \right )^{1/2} \frac{m_4}{m_3} \left ( \frac{1-e_2^2}{1-e_3^2} \right)^{3/2}.
\end{align}
Orbits 1 and 2 are initially mildly mutually inclined, i.e., $i_{12,\mathrm{i}} \simeq 75.0^\circ$. Nevertheless, since orbit 1 is relatively tight ($a_{1,\mathrm{i}}\simeq0.7\,\au$), the maximum eccentricity reached during the secular oscillations is high enough to gradually shrink it, thereby gradually increasing $\mathcal{R}_{3+1}$, which was initially small ($\mathcal{R}_{3+1}\simeq 2\times10^{-3}$). After $\simeq 4500\,\mathrm{Myr}$, orbit 1 shrinks significantly, increasing $\mathcal{R}_{3+1}$ to $\mathcal{R}_{3+1}\sim 10^{-1}$. At the same time, the eccentricity oscillations in orbit 2, which initially had small amplitude, increase significantly, with $e_2$ reaching $e_2\simeq 0.97$. The enhanced eccentricity of orbit 2 has some effect on orbit 1; around $6000\,\mathrm{Myr}$, orbit 1 is slightly excited in eccentricity, but due to tidal evolution it immediately circularizes. This process, which causes several `bumps' in $a_1$ around   $6000\,\mathrm{Myr}$, continues several times until $a_1$ becomes small enough to become completely decoupled from orbit 2.

The enhanced eccentricity of orbit 2 after tidal migration of orbit 1 can be understood by noting that the eccentricity oscillations in orbit 2 due to the secular torque of orbit 3 are initially (partially) quenched due to apsidal precession imposed by orbit 1. This phenomenon is the same as described by \citet{2015MNRAS.449.4221H}. Orbits 2 and 3 are initially inclined by $i_{23,\mathrm{i}} \simeq 81.2^\circ$; therefore, eccentricity oscillations with amplitude $\sim \sqrt{1-5/3 \cos^2 (i_{23,\mathrm{i}})}\simeq 0.98$ are to be expected. However, due to apsidal precession induced by orbit 1, the actual maximum eccentricity reached is $\simeq 0.46$. This is reproduced with the dark blue dotted line in \F\,\ref{fig:example2}, which shows a semianalytic calculation of the maximum eccentricity of orbit 2 using the method based on the secular Hamiltonian described in section 3.4.2 of \citeauthor{2015MNRAS.449.4221H} (\citeyear{2015MNRAS.449.4221H}; the approach is similar to the analytic method to compute the maximum eccentricity in the presence of additional sources of apsidal motion in triple systems, see, e.g., \citealt{2002ApJ...578..775B,2003ApJ...598..419W,2011ApJ...741...82T,2015MNRAS.447..747L,2018MNRAS.481.4907G}). After orbit 1 shrinks, the apsidal precession rate imposed by orbit 1 on orbit 2 decreases, thereby increasing the maximum eccentricity in orbit 2 to $\simeq 0.97$, close to the expected value in the equivalent three-body case. Before and after tidal migration, the semianalytic calculation (dark blue dotted line) is consistent with the numerical solutions to the equations of motion. 

In this example, the enhanced eccentricity of orbit 2 in response to the shrinking of the innermost orbit does not dramatically affect the resulting evolution (although, as noted above, the final semimajor axis of the innermost orbit is somewhat decreased). There are, however, other cases (not shown explicitly here) in which the eccentricity of orbit 2 can become large enough after tidal migration of the innermost orbit to trigger dynamical instability of the system. This shows that dynamical instability in 3+1 systems can be triggered not only by mass loss in evolving systems (e.g., \citealt{2018MNRAS.478..620H}), but also due to tidal evolution during the MS.

\section{Population synthesis set-up}
\label{sect:IC}
In this section, we describe the methodology used to generate the systems for the population synthesis simulations. A summary is given in the third column of Table\,\ref{table:IC}. We sample $N_\mathrm{MC}=10^3$ systems for the 2+2 and 3+1 systems, both with and without the effects of flybys (\S\,\ref{sect:meth:flybys}), and with two different assumptions about the orbital distributions.

\subsection{Masses}
\label{sect:IC:masses}
Our focus is on systems with solar-type MS stars. We set the mass of the primary (most massive) star to $m_1 = 1\,\msun$. The MS time-scale for this mass (and assuming solar metallicity) is $\simeq 10\,\mathrm{Gyr}$, which is also the integration time in our simulations. The secondary mass, $m_2$ ($m_2<m_1$), is sampled from a flat mass ratio distribution, $q_1 = m_2/m_1$ \citep{2012Sci...337..444S,2013ARA&A..51..269D,2017ApJS..230...15M}. From $m_1$ and $m_2$, we sample $m_3$ according to $m_3=(m_1+m_2)q_2$, where $q_2$ has a flat distribution. Lastly, we sample $m_4$ according to $m_4=q_2 m_3$. We reject any sampled combination of masses $m_2$, $m_3$, and $m_4$ if the masses do not satisfy the restrictions $0.1<m_i/\msun<1$ for $i=2,3,4$. We note that this approach implies that the masses of stars 3 and 4 are correlated with those of stars 1 and 2. 

\subsection{Orbits}
\label{sect:IC:orbits}
We adopt two different assumptions about the orbital distributions, in order to establish the sensitivity of our results on the underlying assumed distributions. In case (1), we draw three orbital periods from a Gaussian distribution in $\mathrm{log}_{10}(P_{\mathrm{orb},i}/\uyr)$, with a mean of 5.03 and a standard deviation of 2.28, and $1<\mathrm{log}_{10}(P_{\mathrm{orb},i}/\uyr)<9$ \citep{2010ApJS..190....1R}. The lower limit is approximately the period for which the stars are expected to merge during the pre-MS. The upper limit of $10^{9}\,\uyr$ is 10 times lower than the usual limit of $10^{10}\,\uyr$ in order to better match observed quadruple systems (see \S\,\ref{sect:IC:comp} below). The corresponding semimajor axes are computed according to the configuration using Kepler's law. In addition, three eccentricities are drawn from a Rayleigh distribution between 0.01 and 0.95 with an rms width of 0.33 \citep{2010ApJS..190....1R}. In case (2), we adopt \"{O}pic's law \citep{opic1924}, i.e., flat distributions in $\log_{10}(a_i)$, subject to the same orbital period ranges, i.e,. $1<\mathrm{log}_{10}(P_{\mathrm{orb},i}/\uyr)<9$. The eccentricity distributions are assumed to be flat in this case, again with $0.01<e_i<0.95$. 

In both cases, we impose stability criteria to ensure that the systems are dynamically stable using the criterion of \citeauthor{2001MNRAS.321..398M} (\citeyear{2001MNRAS.321..398M}; implicitly assuming that this criterion also applies to quadruple systems). Note that \citet{2018MNRAS.474...20H} recently reinvestigated the criterion of \citet{2001MNRAS.321..398M}, and found that it predicts stability against ejections reasonably well for a wide range of parameters. We also impose conditions to ensure that the inner orbits would not evolve due to tides in the absence of secular evolution (i.e., as isolated binaries). Specifically, let the criterion of \citet{2001MNRAS.321..398M}, as applied to a triple system with the inner and outer orbits indicated with `in' and `out', respectively, be denoted with $a_\mathrm{out}/a_\mathrm{in} > f_\mathrm{MA01}(m_\mathrm{in},m_\mathrm{out},e_\mathrm{out})$, where the latter function is given by
\begin{align}
f_\mathrm{MA01}(m_\mathrm{in},m_\mathrm{out},e_\mathrm{out}) \equiv \frac{2.8}{1-e_\mathrm{out}} \left [ \left (1+\frac{m_\mathrm{out}}{m_\mathrm{in}} \right ) \frac{1+e_\mathrm{out}}{\sqrt{1-e_\mathrm{out}}} \right ]^{2/5}.
\end{align}
For 2+2 systems, we impose
\begin{subequations}
\begin{align}
a_2/a_1 &>  f_\mathrm{MA01}(m_1+m_2,m_3+m_4,e_3); \\
a_3/a_2 &> f_\mathrm{MA01}(m_3+m_4,m_1+m_2,e_3); \\
a_1\left(1-e_1^2\right) &> a_\mathrm{crit}; \\
a_2\left(1-e_2^2\right) &> a_\mathrm{crit},
\end{align}
\end{subequations}
whereas for 3+1 systems, we require
\begin{subequations}
\begin{align}
a_2/a_1 &>  f_\mathrm{MA01}(m_1+m_2,m_3,e_2); \\
a_3/a_2 &> f_\mathrm{MA01}(m_1+m_2+m_3,m_4,e_3); \\
a_1\left(1-e_1^2\right) &> a_\mathrm{crit}.
\end{align}
\end{subequations}
Here, $a_\mathrm{crit}$ is the semimajor axis corresponding to the smallest allowed orbital period of 10 d. The restrictions on the semilatus recti, $a_i\left(1-e_i^2\right)$, ensure that the inner orbits do not shrink to an orbital period less than 10 d due to tides. This implies that any shrinkage of the inner orbits to less than 10 d observed in the simulations can be fully ascribed to secular evolution (i.e., tides triggered by enhanced eccentricity due to secular evolution). 

The initial orbital orientations for all configurations are assumed to be random, i.e., for each orbit $i$, flat distributions are assumed for $\cos(i_i)$, $\omega_i$ and $\Omega_i$, where $i_i$, $\omega_i$ and $\Omega_i$ are the inclination, argument of periapsis, and longitude of the ascending node, respectively, of orbit $i$ (the orbital elements are defined with respect to an arbitrary fixed frame).

\begin{figure}
\center
\includegraphics[scale = 0.45, trim = 10mm 10mm 0mm 10mm]{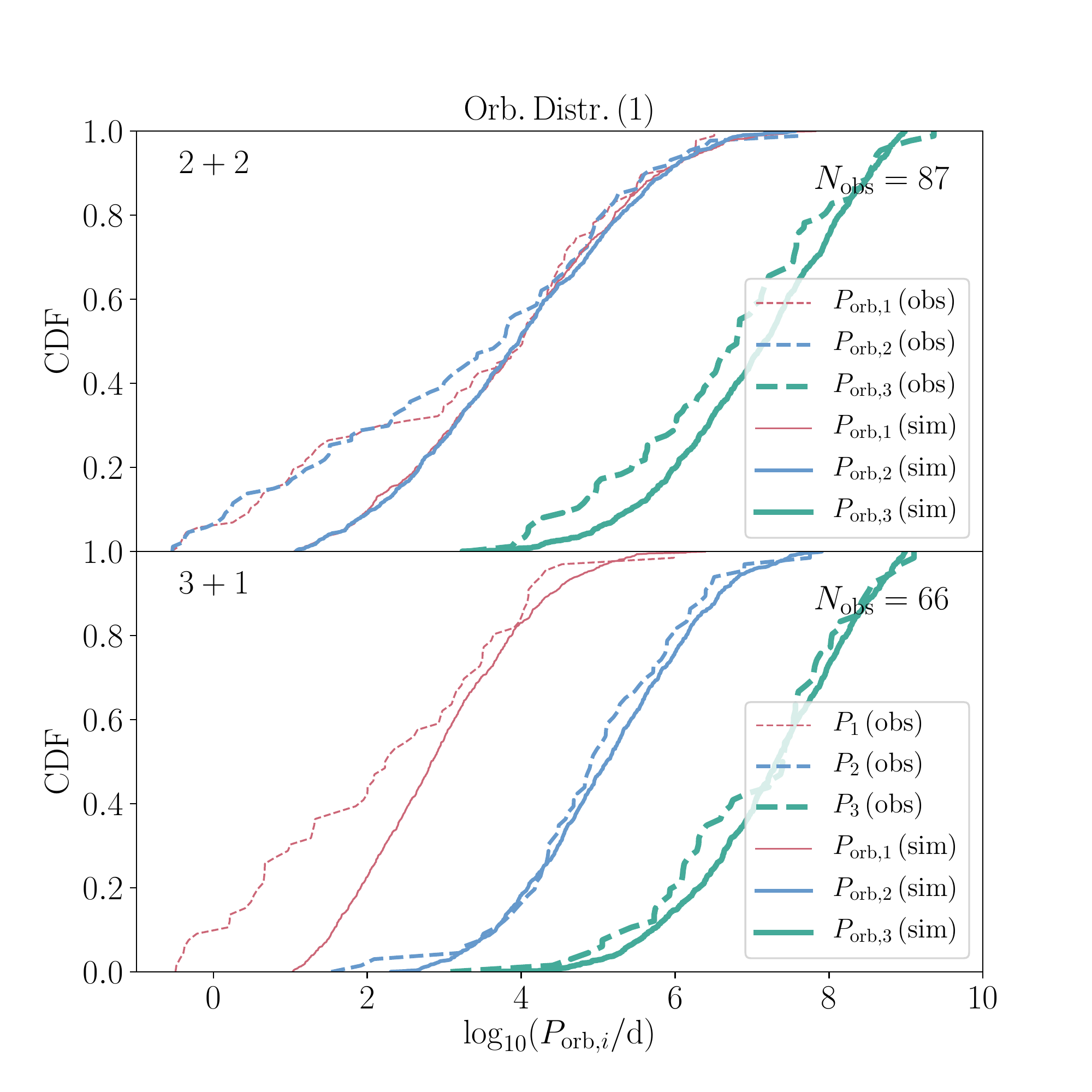}
\caption { Cumulative distribution comparison of the (initial) orbital periods sampled from the case 1 sampling method described in \S\,\ref{sect:IC} (Gaussian distributions in the logarithmic periods and Rayleigh eccentricity distributions; solid lines) to observational data satisfying similar requirements from the MSC (\citealt{1997A&AS..124...75T,2018ApJS..235....6T}; dashed lines). The top (bottom) panel corresponds to the 2+2 (3+1) configuration. Orbits are indicated with different colours: red, blue, and green for orbits 1 through 3, and with increasing line widths. The number of systems in the MSC is indicated in the top right of each panel.  }
\label{fig:IC_comp_per1}
\end{figure}

\begin{figure}
\center
\includegraphics[scale = 0.45, trim = 10mm 10mm 0mm 10mm]{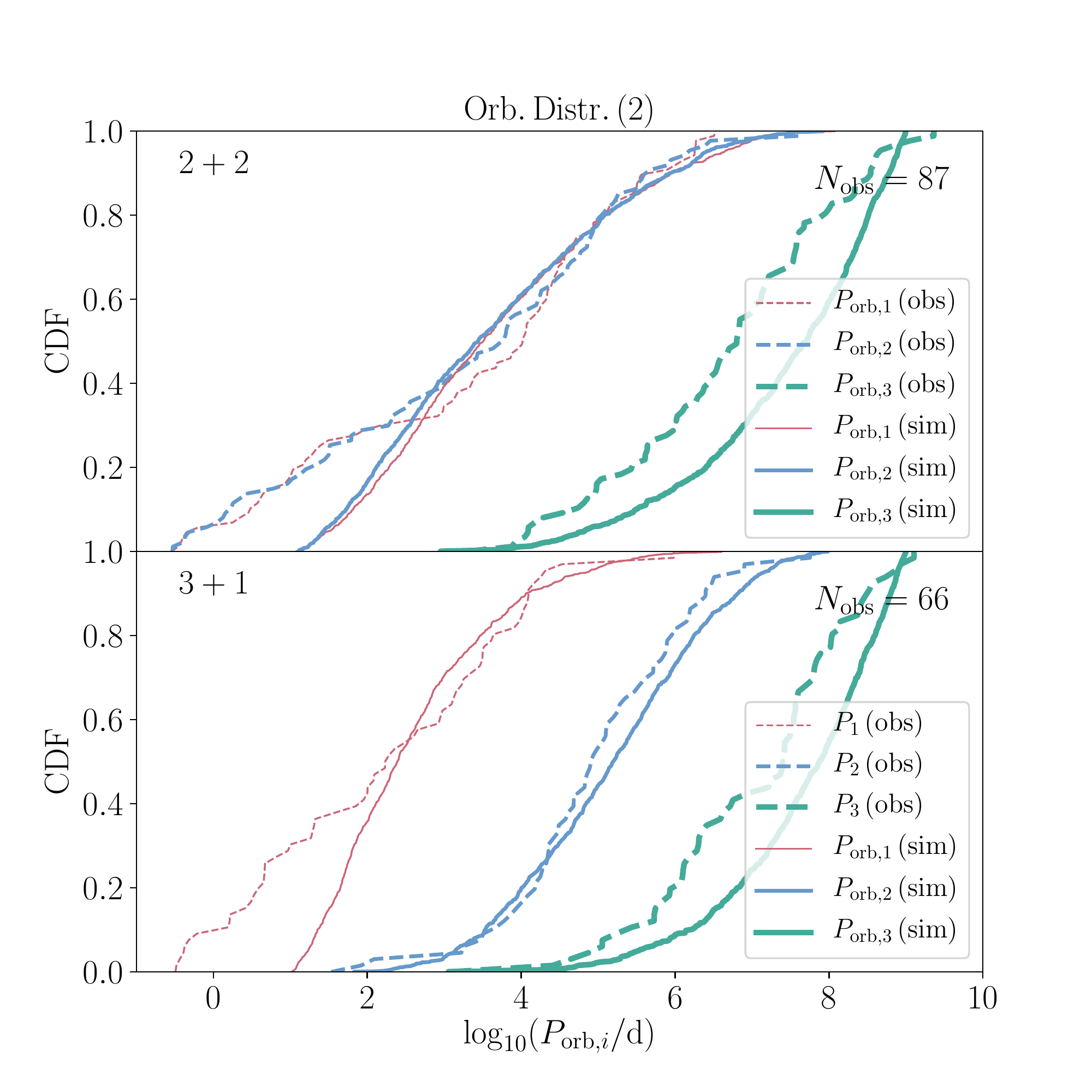}
\caption { Similar to \F\,\ref{fig:IC_comp_per1}, now for the case 2 sampling method (flat distributions in $\log_{10}a_i$ and $e_i$). }
\label{fig:IC_comp_per2}
\end{figure}

\begin{figure}
\center
\includegraphics[scale = 0.45, trim = 10mm 10mm 0mm 10mm]{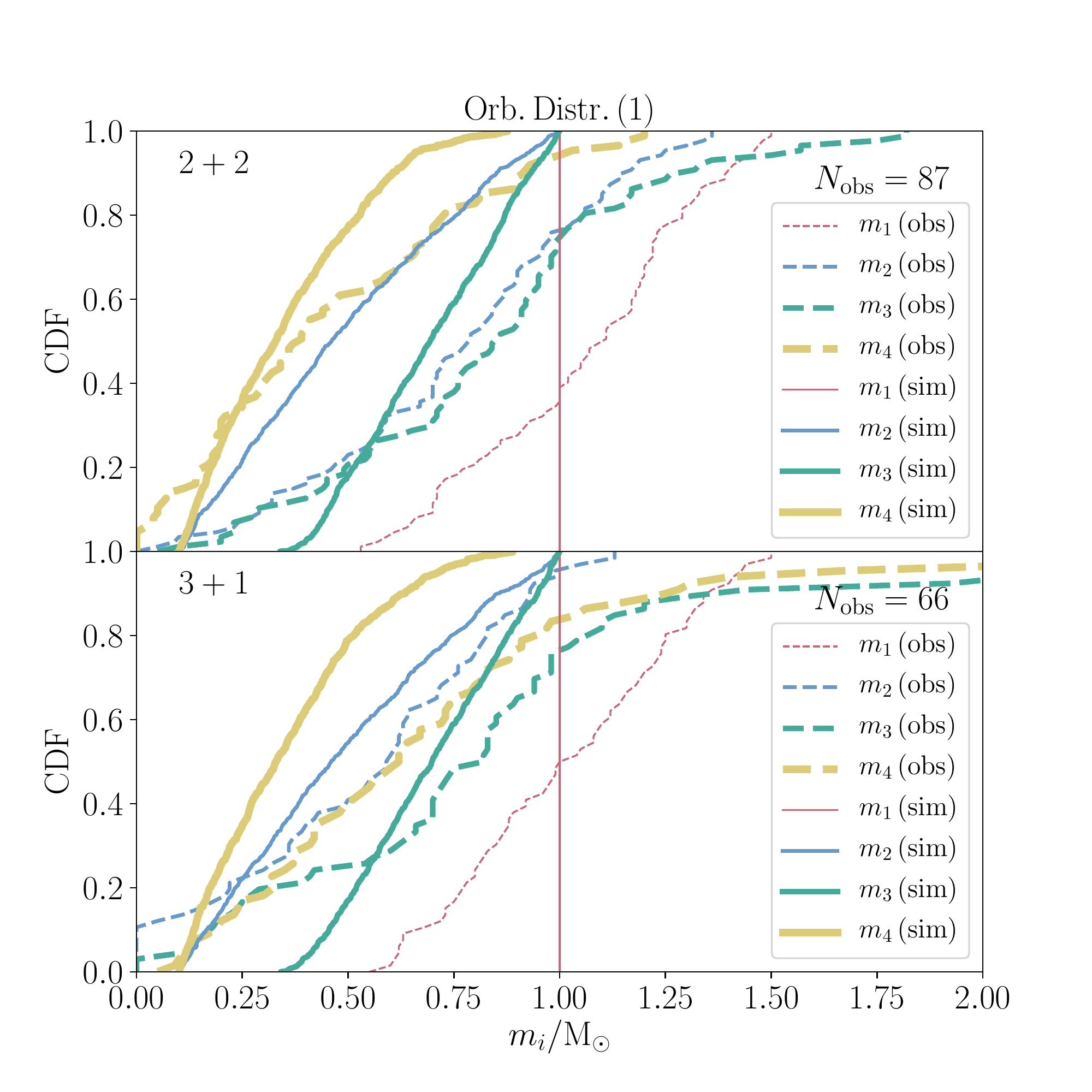}
\caption { Similar to \F\,\ref{fig:IC_comp_per1}, here comparing the distributions of the four masses. Note that the mass distributions are the same for the two adopted orbital sampling methods.}
\label{fig:IC_comp_m}
\end{figure}

\subsection{Comparison to the Multiple Star Catalogue}
\label{sect:IC:comp}
The statistics of orbital properties of quadruple star systems are much less constrained than those of isolated binary (and triple) stars. Nevertheless, here we briefly compare the distributions of sampled systems to observations. We take data from the Multiple Star Catalogue (MSC; \citealt{1997A&AS..124...75T,2018ApJS..235....6T}), selecting systems with primary masses $0.5<m_1/\msun<1.5$ and all orbits known, giving 87 (66) systems for the 2+2 (3+1) configuration. We compare our sampling methodology to the MSC in terms of the orbital period distributions (Figs.\,\ref{fig:IC_comp_per1} and \ref{fig:IC_comp_per2}), and the mass distributions (\F\,\ref{fig:IC_comp_m}). In these figures, the top (bottom) panel corresponds to the 2+2 (3+1) configuration. 

The first orbital sampling method (lognormal orbital period and Rayleigh eccentricity distributions; see \F\,\ref{fig:IC_comp_per1}) agrees reasonably with the MSC for periods $\gtrsim 10^3\,\ud$. There is a clear excess of systems with inner periods between $\sim 1$ and $\sim 10^2\,\ud$, as noted before by \citet{2008MNRAS.389..925T}. Evidently, our aim in \S\,\ref{sect:pop_syn} is to establish whether tidal migration coupled with secular evolution can reproduce such an excess of short-period systems. The second sampling method (flat distributions in $\log_{10} a_i$ and the eccentricities; see \F\,\ref{fig:IC_comp_per2}) compares less favourably to the MSC at periods $\gtrsim 10^3\,\ud$, especially for the outermost orbits. 
 
The median $m_1$ in the extracted sample of the MSC agrees with the assumed $m_1=1\,\msun$ (see \F\,\ref{fig:IC_comp_m}). The other observed masses, $m_2$, $m_3$ and $m_4$, tend to be somewhat larger than the sampled masses. However, we do not anticipate a very strong dependence of our results on the masses.

\section{Population synthesis results}
\label{sect:pop_syn}
Here, we present the results from the population synthesis simulations. The initial conditions were discussed in \S\,\ref{sect:IC}. We first focus on the occurrences of tidal migration and other outcomes (\S\,\ref{sect:pop_syn:frac}). We then consider the inner orbital period distributions mediated by tidal and secular evolution (\S\,\ref{sect:pop_syn:in}). Other orbital properties are considered in \S\,\ref{sect:pop_syn:orb}, and in \S\,\ref{sect:pop_syn:time} we discuss the migration times. 

\subsection{Outcome fractions}
\label{sect:pop_syn:frac}

\begin{table*}
\scriptsize
\begin{tabular}{lcccccccccccccccc}
\toprule
& \multicolumn{16}{c}{Fraction} \\
& \multicolumn{8}{c}{$2+2$} &\multicolumn{8}{c}{$3+1$} \\
& \multicolumn{4}{c}{$t_x$} & \multicolumn{4}{c}{10 Gyr} & \multicolumn{4}{c}{$t_x$} & \multicolumn{4}{c}{10 Gyr} \\
& \multicolumn{2}{c}{Orb. Distr. (1)} & \multicolumn{2}{c}{Orb. Distr. (2)} & \multicolumn{2}{c}{Orb. Distr. (1)} & \multicolumn{2}{c}{Orb. Distr. (2)} & \multicolumn{2}{c}{Orb. Distr. (1)} & \multicolumn{2}{c}{Orb. Distr. (2)} & \multicolumn{2}{c}{Orb. Distr. (1)} & \multicolumn{2}{c}{Orb. Distr. (2)}\\
& NF & F & NF & F & NF & F & NF & F & NF & F & NF & F & NF & F & NF & F \\
\midrule
$\mathrm{No\,Mig.}$ & 0.790 & 0.772 & 0.842 & 0.805 & 0.763 & 0.727 & 0.822 & 0.764 & 0.507 & 0.576 & 0.577 & 0.590 & 0.420 & 0.535 & 0.521 & 0.528  \\
$\mathrm{Mig. \,O1}$ & 0.015 & 0.019 & 0.034 & 0.041 & 0.019 & 0.025 & 0.034 & 0.041 & 0.093 & 0.084 & 0.131 & 0.122 & 0.104 & 0.086 & 0.144 & 0.131  \\
$\mathrm{Mig. \,O2}$ & 0.026 & 0.032 & 0.025 & 0.024 & 0.028 & 0.038 & 0.030 & 0.027 & --- & --- & --- & --- & --- & --- & --- & --- \\
$\mathrm{Mig. \,O1+O2}$ & 0.002 & 0.001 & 0.004 & 0.004 & 0.002 & 0.003 & 0.004 & 0.006 & --- & --- & --- & --- & --- & --- & --- & --- \\
$\mathrm{RLOF\,O1}$ & 0.055 & 0.063 & 0.033 & 0.043 & 0.059 & 0.073 & 0.038 & 0.056 & 0.141 & 0.142 & 0.117 & 0.119 & 0.152 & 0.151 & 0.127 & 0.131  \\
$\mathrm{RLOF\,O2}$ & 0.106 & 0.098 & 0.053 & 0.053 & 0.115 & 0.109 & 0.057 & 0.061 & --- & --- & --- & --- & --- & --- & --- & --- \\
$\mathrm{Dyn.\, Inst.\,O1}$ & 0.001 & 0.006 & 0.003 & 0.018 & 0.001 & 0.012 & 0.004 & 0.022 & 0.127 & 0.138 & 0.100 & 0.105 & 0.138 & 0.142 & 0.102 & 0.118  \\
$\mathrm{Dyn.\, Inst.\,O2}$ & 0.002 & 0.007 & 0.002 & 0.011 & 0.002 & 0.010 & 0.002 & 0.021 & 0.027 & 0.027 & 0.018 & 0.039 & 0.029 & 0.040 & 0.019 & 0.056  \\
$\mathrm{Unbound\,Flyby}$ & --- & 0.001 & --- & 0.001 & --- & 0.001 & --- & 0.001 & --- & 0.000 & --- & 0.000 & --- & 0.000 & --- & 0.000  \\
$\mathrm{Time\,exceeded}$ & 0.003 & 0.001 & 0.004 & 0.000 & 0.011 & 0.002 & 0.009 & 0.001 & 0.105 & 0.033 & 0.057 & 0.025 & 0.157 & 0.046 & 0.087 & 0.036  \\
\bottomrule
\end{tabular}
\caption{ Fractions of outcomes from the population synthesis calculations. Columns 2-9: 2+2 systems; columns 10-17: 3+1 systems. Fractions are shown after a random time (indicated with the columns $t_x$) or after 10 Gyr of evolution (if applicable; otherwise, the end time is the stopping condition time). Results are shown for the first (`Orb. Distr. (1)') and second (`Orb. Distr. (2)') assumptions about the orbital distributions (see \S\,\ref{sect:IC:orbits}), and without the inclusion of flybys (`NF'), or with inclusion (`F'). The fractions in each column are obtained from $N_\mathrm{MC}=10^3$ simulations, and are rounded to three decimal places. Each row corresponds to a different outcome; see the text in \S\,\ref{sect:pop_syn:frac} for a description. Note that some outcomes do not apply in the simulations (indicated with `---'): unbound systems due to flybys in the `NF' cases, and, for the 3+1 configuration, migration in orbit 2, migration in both orbits 1 and 2, and RLOF in orbit 2.}
\label{table:fractions}
\end{table*}

In Table\,\ref{table:fractions}, we show the fractions of various outcomes in the simulations. We distinguish between the following outcomes, which are partially based on the stopping conditions (\S\,\ref{sect:meth:sc}).
\begin{itemize}
\item Migration of an inner orbit, i.e., $P_{\mathrm{orb},i}<10\,\ud$, where $i$ can be 1 or 2 for the 2+2 configuration, or 1 for the 3+1 configuration.
\item RLOF in orbit 1 or 2 (2+2 configuration), or orbit 1 only (3+1 configuration).
\item Dynamical instability in orbit 1 or 2 with respect to their parent according to the criterion of \citet{2001MNRAS.321..398M}. Note that, in the 2+2 configuration, the parent to orbits 1 and 2 is orbit 3, whereas for the 3+1 configuration the parent of orbit 1 is orbit 2, and the parent of orbit 2 is orbit 3.
\item One or more of the orbits in the quadruple system (typically, the outer orbit) become unbound due to an impulsive encounter. This is a rare occurrence, and evidently only applies if $\renc>0$.
\item The integration time exceeds the set CPU wall time of 24 hr (\S\,\ref{sect:meth:sc}). These systems potentially present an uncertainty in the results. However, we show below in \S\,\ref{sect:discussion:exceed} that this likely does not affect the final orbital period distributions in the simulations.
\item In all other cases, we mark the outcome as `no migration'.
\end{itemize}
For each outcome, we show the corresponding fraction, based on the $N_\mathrm{MC}=10^3$ simulations, for the 2+2 and 3+1 configurations, and various sets of simulations: with the first (`Orb. Distr. (1)') and second (`Orb. Distr. (2)') assumptions about the orbital distributions (see \S\,\ref{sect:IC:orbits}), and with (`F') and without (`NF') the inclusion of flybys. Also, we show fractions after 10 Gyr of evolution (if applicable; otherwise, the end time is the stopping condition time), or after a random time between 0 and 10 Gyr ($t_x$), appropriate for continuous star formation. 

The most likely outcome in all cases is `no migration'. Note that the eccentricities may still have been excited (see \S\,\ref{sect:pop_syn:orb:e} below). In a few per cent for the 2+2 systems, and in up to $\sim 14\%$ for the 3+1 systems, tidal migration occurs in orbit 1 or 2. In the 2+2 configuration, double migration, of which we showed an example in \S\,\ref{sect:examples:1}, occurs in a few tenths of per cents, i.e., is relatively rare. RLOF is triggered in up to $\sim 11\%$ for the 2+2 systems, and $\sim 15\%$ for the 3+1 systems. In the 2+2 systems, dynamical instability rarely occurs, which can be understood by noting that dynamical instability can only be triggered in our simulations by flybys, or, more rarely, if the outer orbit eccentricity, $e_3$, is enhanced, which can only occur due to high-order terms (higher than quadrupole-order), and therefore only in compact systems. For the 3+1 systems, dynamical instability is relatively common for the orbit 1-2 pair (up to $\sim 14\%$). This can be attributed to enhanced eccentricity of orbit 2 if it is inclined relative to orbit 3. As noted above, the unbinding of the system due to flybys rarely occurs. The run time is exceeded mostly for 3+1 systems, and this is further addressed in \S\,\ref{sect:discussion:exceed}.

There are typically no major differences between the results for the two assumptions about the orbital distributions (Orb. Distr. (1)' and `Orb. Distr. (2)'). Typically, the second assumed orbital distributions (flat distributions in $\log_{10}a_i$ and $e_i$) tend to lead to fewer stronger interactions (the `no migration' fractions are larger, and all other fractions are lower). This can be attributed to the generally wider orbits compared to the first assumed orbital distributions (see \S\,\ref{sect:IC:orbits}). As can be natively expected, flybys tend to increase the occurrence of strong interactions (decreasing the no-migration fractions, and increasing the others).

\subsection{Inner orbital period distributions}
\label{sect:pop_syn:in}

\begin{figure*}
\center
\includegraphics[scale = 0.45, trim = 10mm -2mm 0mm 10mm]{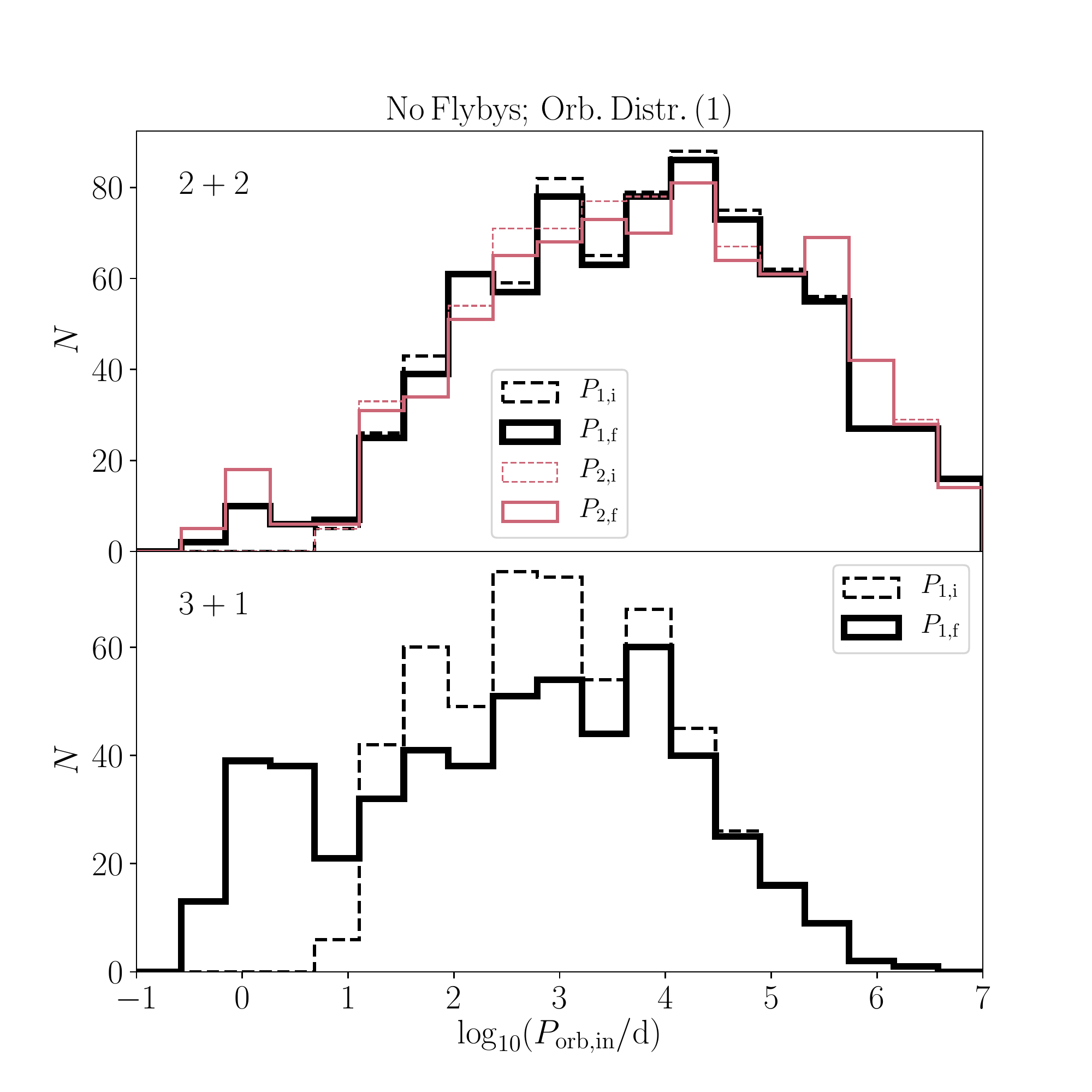}
\includegraphics[scale = 0.45, trim = 10mm -2mm 0mm 10mm]{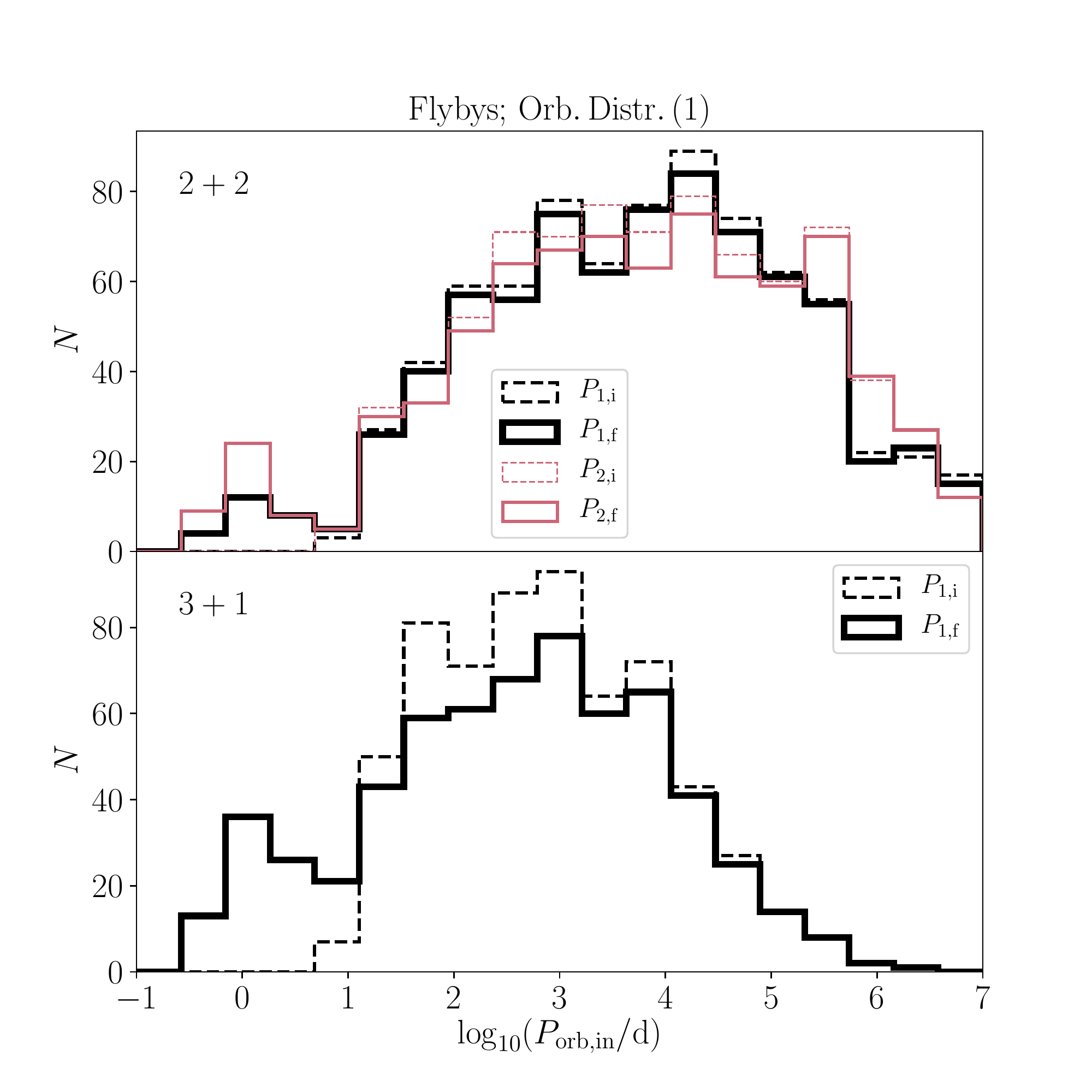}
\includegraphics[scale = 0.45, trim = 10mm 10mm 0mm 10mm]{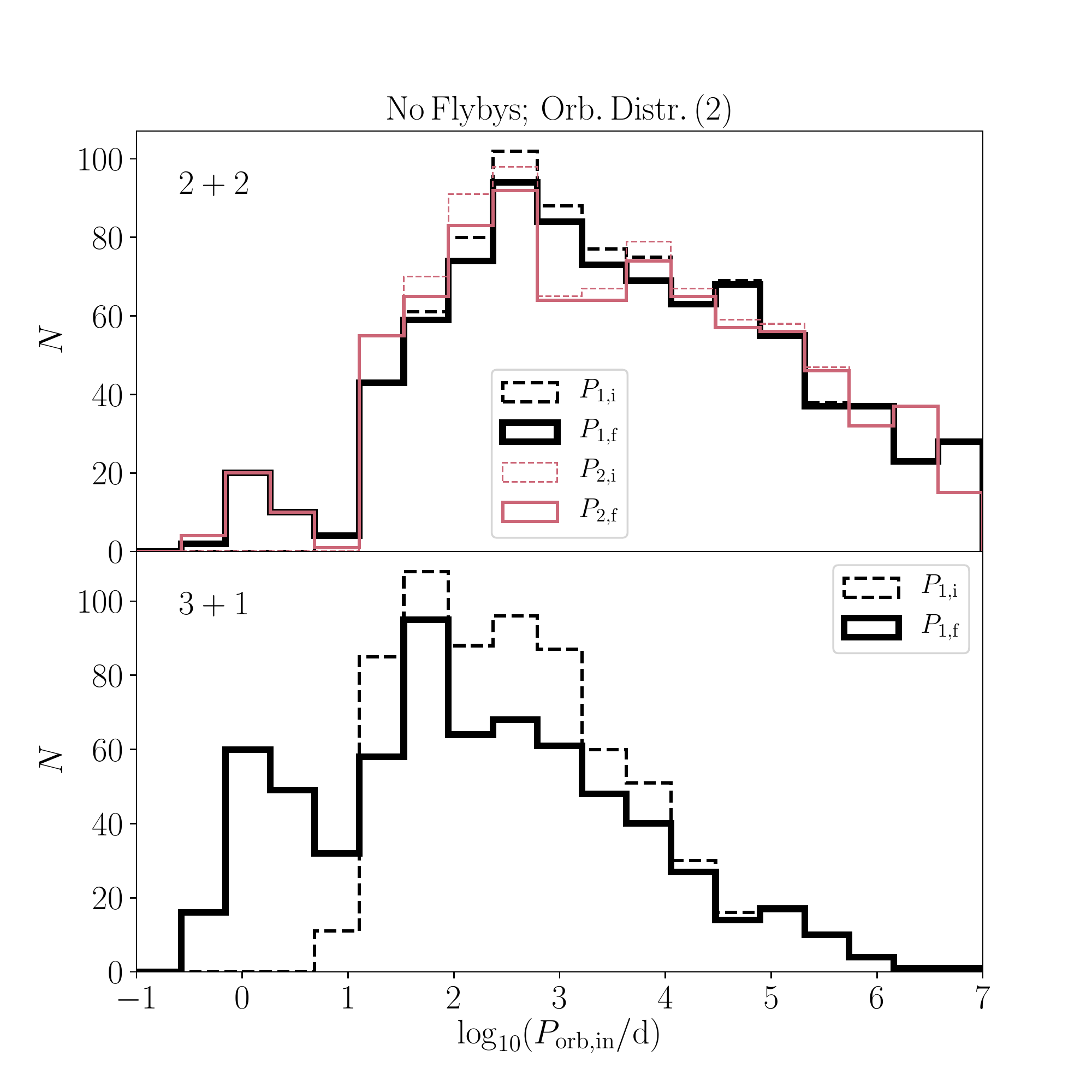}
\includegraphics[scale = 0.45, trim = 10mm 10mm 0mm 10mm]{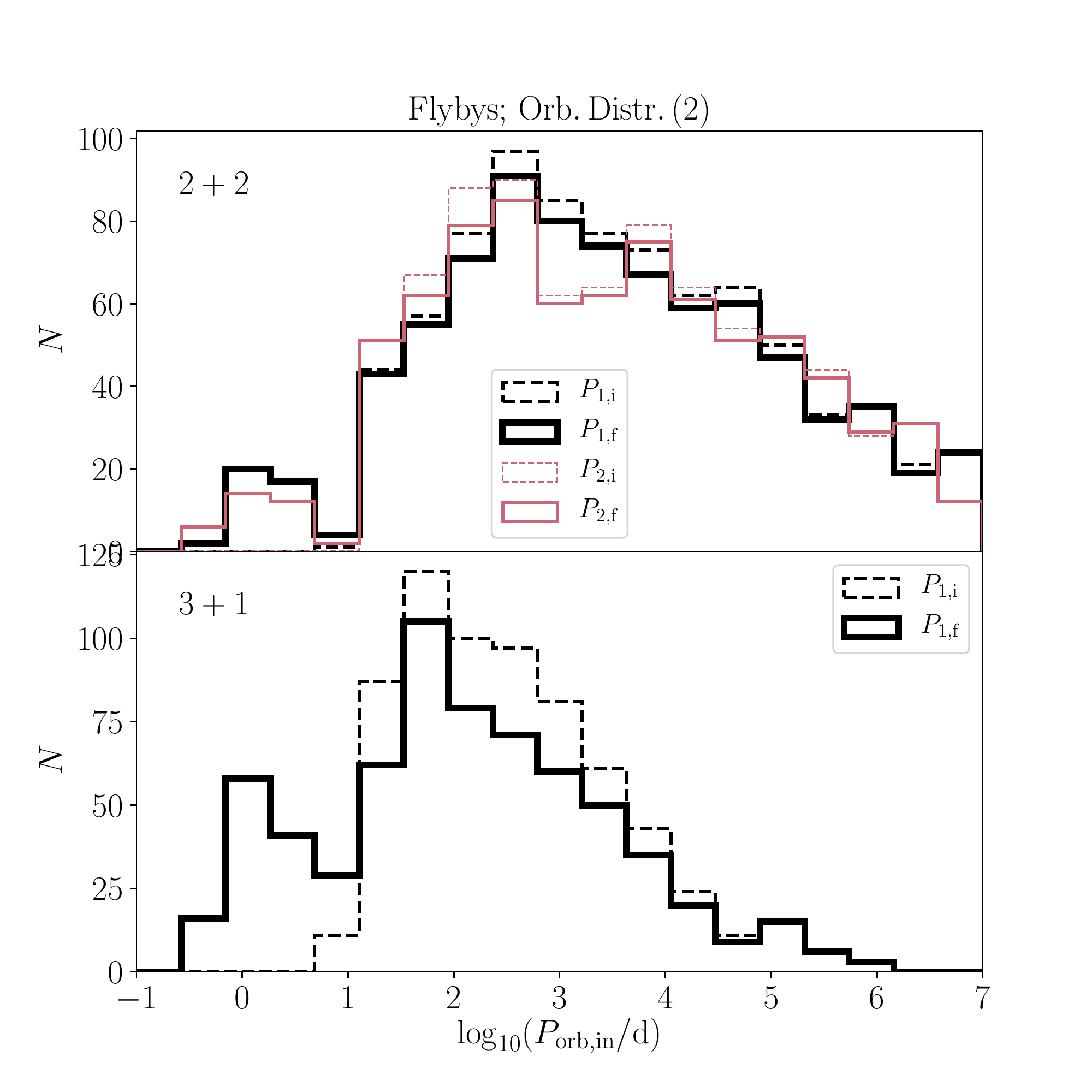}
\caption { Histograms of the inner orbital period distributions in the simulations. For the 2+2 configuration, a distinction is made between orbits 1 (thicker black lines), and 2 (thinner red lines). Initial (final) distributions are shown with dashed (solid) lines. The four sets of panels (each set containing two panels for the two configurations) correspond to whether or not flybys were included, and which orbital distribution was assumed, indicated above the top panel. }
\label{fig:pop_syn_p_ins}
\end{figure*}

In \F\,\ref{fig:pop_syn_p_ins}, we show histograms of the inner orbital period distributions in the simulations. The data shown correspond to the `no migration' and `migration' outcomes of Table\,\ref{table:fractions}. For the 2+2 configuration, we make a distinction between orbits 1 (thicker black lines), and 2 (thinner red lines). Initial (final) distributions are shown with dashed (solid) lines. The four sets of panels (each set containing two panels for the two configurations) correspond to whether or not flybys were included, and which orbital distribution was assumed, indicated above the top panel. 

We first note that the inner orbital periods for the 3+1 configuration tend to be shorter compared to the 2+2 configuration, although the underlying distributions for all orbits were assumed to be the same. This can be ascribed to the requirement of dynamical stability and the hierarchy of the system: for 2+2 systems, only two tiers (or levels) of orbits need to fit within $\sim 10$ decades of orbital period, whereas for the 3+1 configuration, three tiers need to fit within the same orbital period range. This was illustrated in Figs.\,\ref{fig:IC_comp_per1} and \ref{fig:IC_comp_per2}, which also show that this property is consistent with observed quadruple systems. 

For both configurations, the initial orbital period distributions were cut off at 10 d; evidently, due to tidal and secular evolution, the distributions develop tails at periods below 10 d, similar to previous Monte Carlo studies of tidal migration in triple systems \citep{2007ApJ...669.1298F,2009ApJ...697.1048P,2014ApJ...793..137N}. These tails are more prominent for the 3+1 systems compared to the 2+2 systems, which is also reflected by the higher migration fractions for the 3+1 configuration in Table\,\ref{table:fractions}. This can be attributed to two effects: (1) typically tighter initial inner orbits for the 3+1 systems, due to the reasons given above; (2) typically stronger secular evolution (i.e., higher eccentricities are reached and therefore tidal migration is more efficient) for the innermost orbit in the 3+1 system, due to the possibility of enhanced outer orbit eccentricity (orbit 2 for the 3+1 configuration). The enhanced secular evolution in the 3+1 systems is also corroborated by the significantly higher RLOF fractions for these systems (see Table\,\ref{table:fractions}).

We note that there are no major qualitative differences in the final orbital period distributions when different assumptions are made about the orbital distributions, and whether or not flybys are included. 

\begin{figure*}
\center
\includegraphics[scale = 0.45, trim = 10mm -2mm 0mm 10mm]{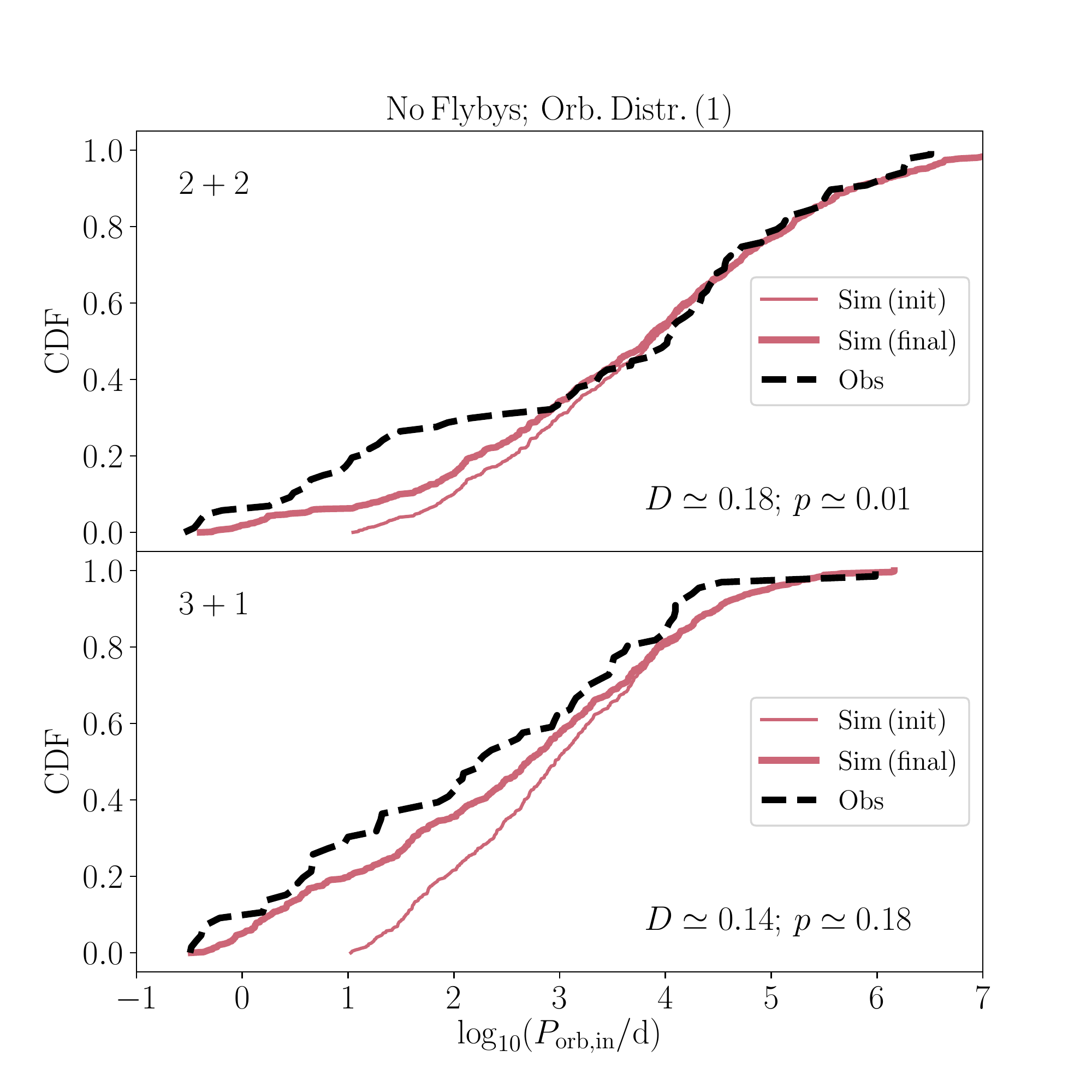}
\includegraphics[scale = 0.45, trim = 10mm -2mm 0mm 10mm]{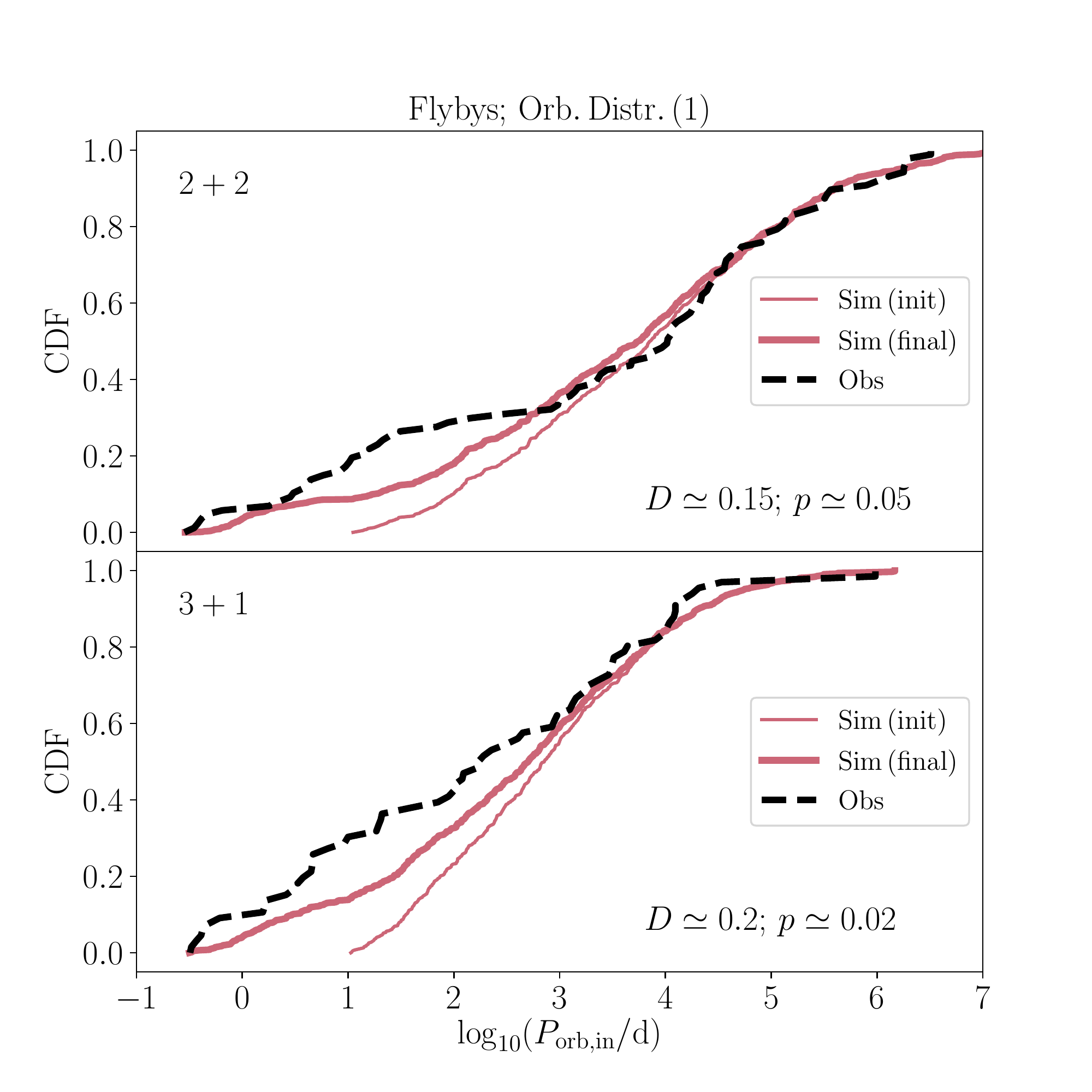}
\includegraphics[scale = 0.45, trim = 10mm 10mm 0mm 10mm]{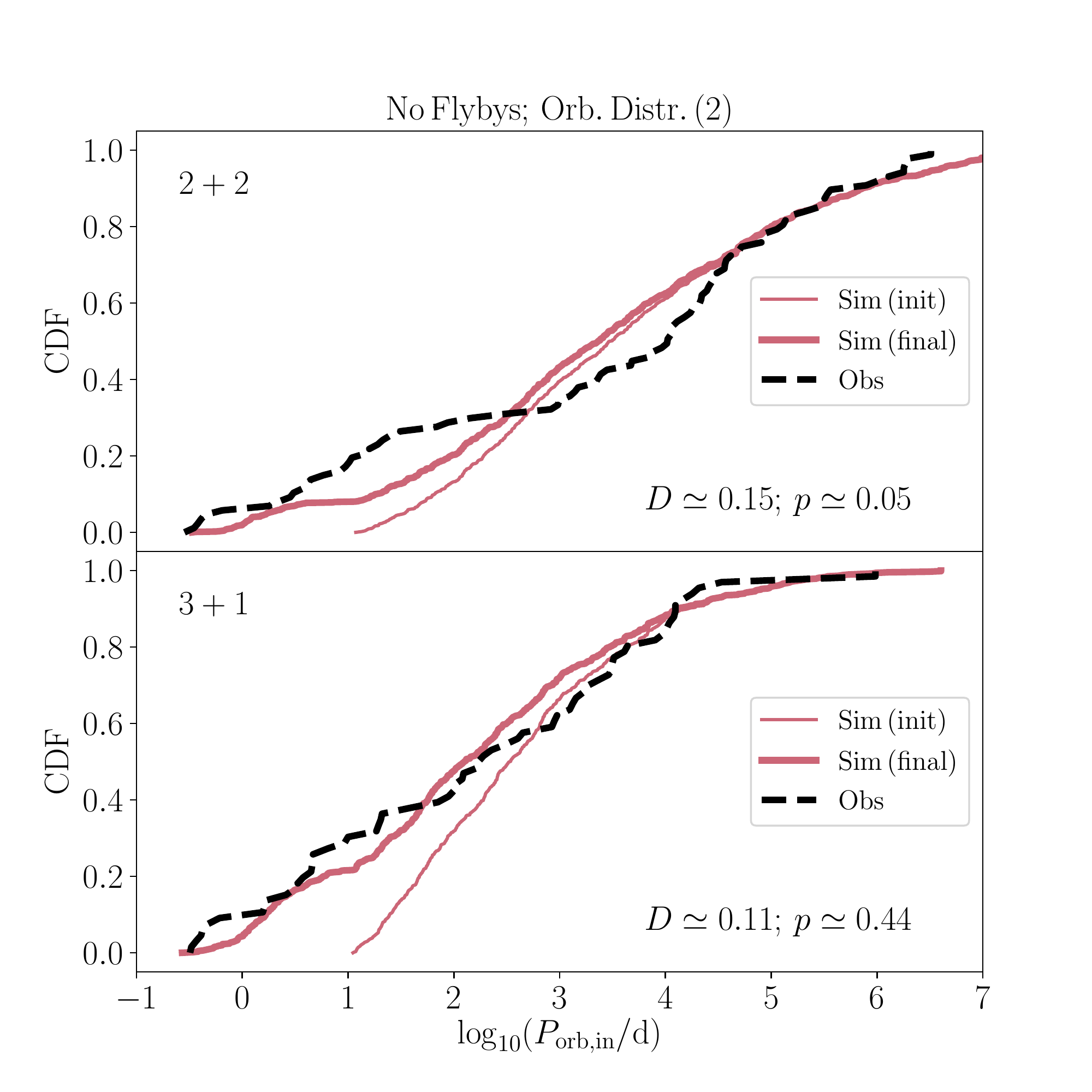}
\includegraphics[scale = 0.45, trim = 10mm 10mm 0mm 10mm]{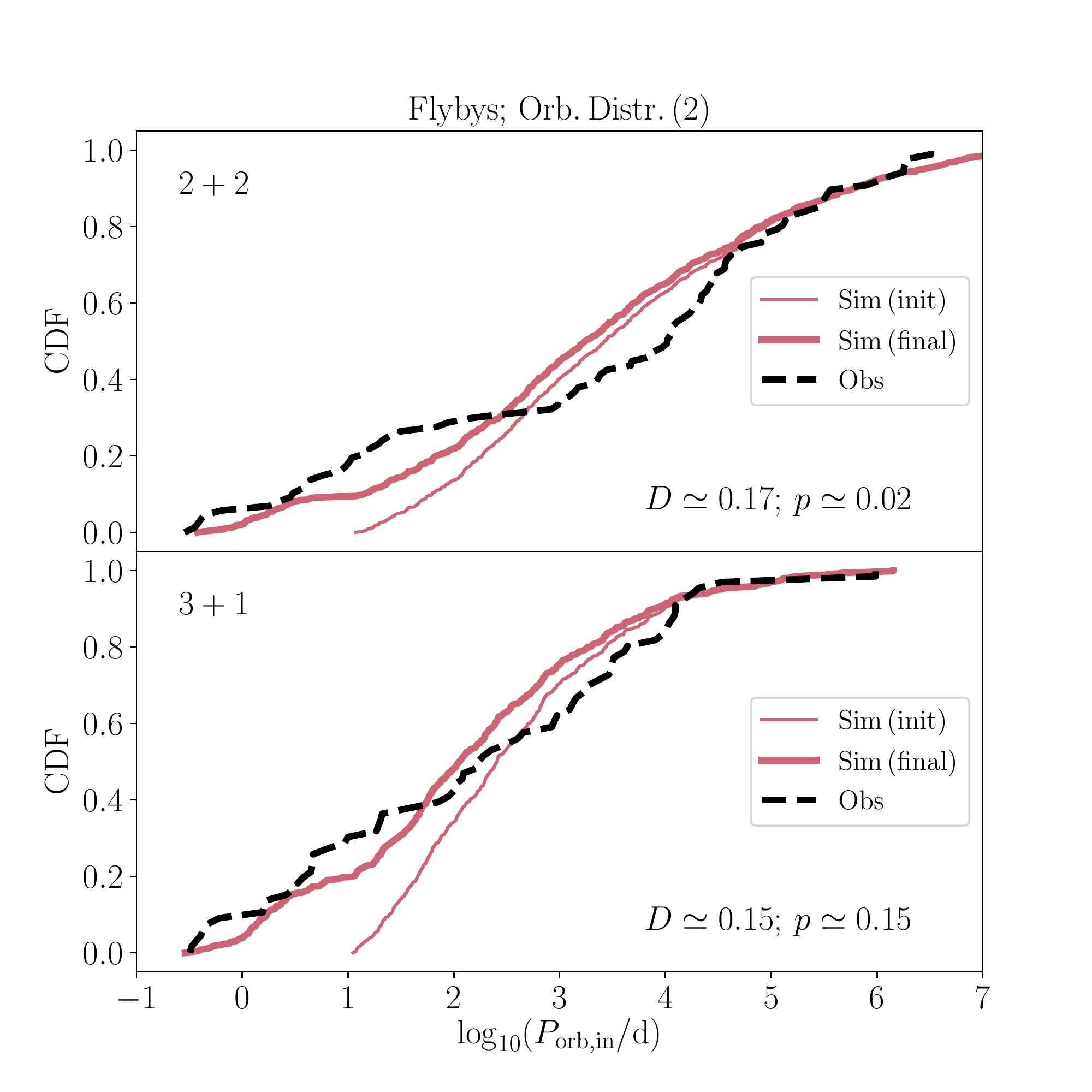}
\caption { Comparison between the simulated (red lines; thin and thick lines correspond to the initial and final distributions, respectively) and observed (black dashed lines) cumulative inner orbital period distributions of solar-type MS quadruple stars. The four sets of panels (each set containing two panels for the two configurations) correspond to whether or not flybys were included, and which orbital distribution was assumed, indicated in the top panel. The observed distributions are adopted from the MSC (\citealt{1997A&AS..124...75T,2018ApJS..235....6T}; see also \S\,\ref{sect:IC:comp}). For the 2+2 configuration, data are combined from orbits 1 and 2. In each panel, the $D$ and $p$ statistics from a K-S test between the final simulated data and the observations are shown. }
\label{fig:pop_syn_p_ins_obs}
\end{figure*}

In \F\,\ref{fig:pop_syn_p_ins_obs}, we show the same data as shown in \F\,\ref{fig:pop_syn_p_ins}, but now comparing, in terms of the cumulative inner orbital period distributions, the data from the simulations (red lines; with the thin and thick lines corresponding to the initial and final distributions, respectively) to the observations (the MSC, \citealt{1997A&AS..124...75T,2018ApJS..235....6T}, with black dashed lines; see also \S\,\ref{sect:IC:comp}). Here, we combine the data of orbits 1 and 2 for the 2+2 configuration. We compare the simulated final orbital periods to the observations using two-sided Kolmogorov-Smirnov (K-S) tests \citep{Kolmogorov_33,smirnov_48}; the $D$ and $p$ statistics are indicated in each panel. 

From \F\,\ref{fig:pop_syn_p_ins_obs}, we observe that, although the simulated distributions are enhanced at short periods ($<10\,\ud$) due to secular and tidal evolution, the observed distributions are more peaked at short periods, in particular for the 2+2 systems. For the latter, the observed distribution contains a sharp rise in systems around 10 d, which is not reproduced in the simulations. Such a sharp rise is less prominent in the observed distribution for 3+1 systems, and the match with the simulations is notably better. Quantitatively, the $p$ values for the 2+2 configuration are $\lesssim 0.05$, whereas for the 3+1 configuration, the $p$ values are generally larger, up to $\simeq 0.44$, although in the case of orbital distribution assumption (1) and flybys included, the $p$ value is low, $p\simeq0.02$. The best statistical agreement with the observations is reached for the 3+1 configuration with no flybys and the second assumption about the orbital distributions (flat in $\log_{10} a_i$ and $e_i$). We further discuss the implications of this result in \S\,\ref{sect:discussion:per}.

\begin{figure}
\center
\includegraphics[scale = 0.45, trim = 10mm -5mm 0mm 10mm]{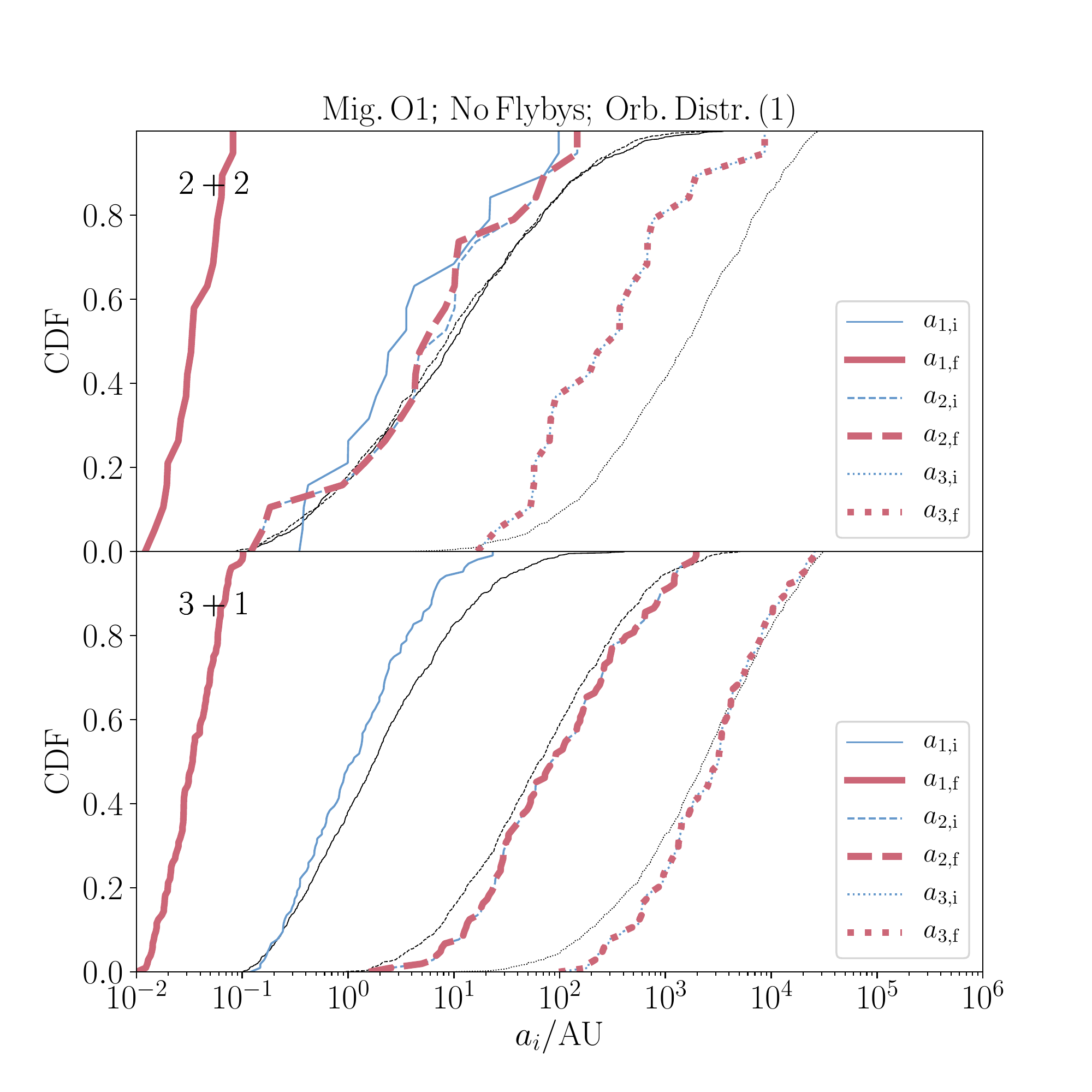}
\includegraphics[scale = 0.45, trim = 10mm 10mm 0mm 10mm]{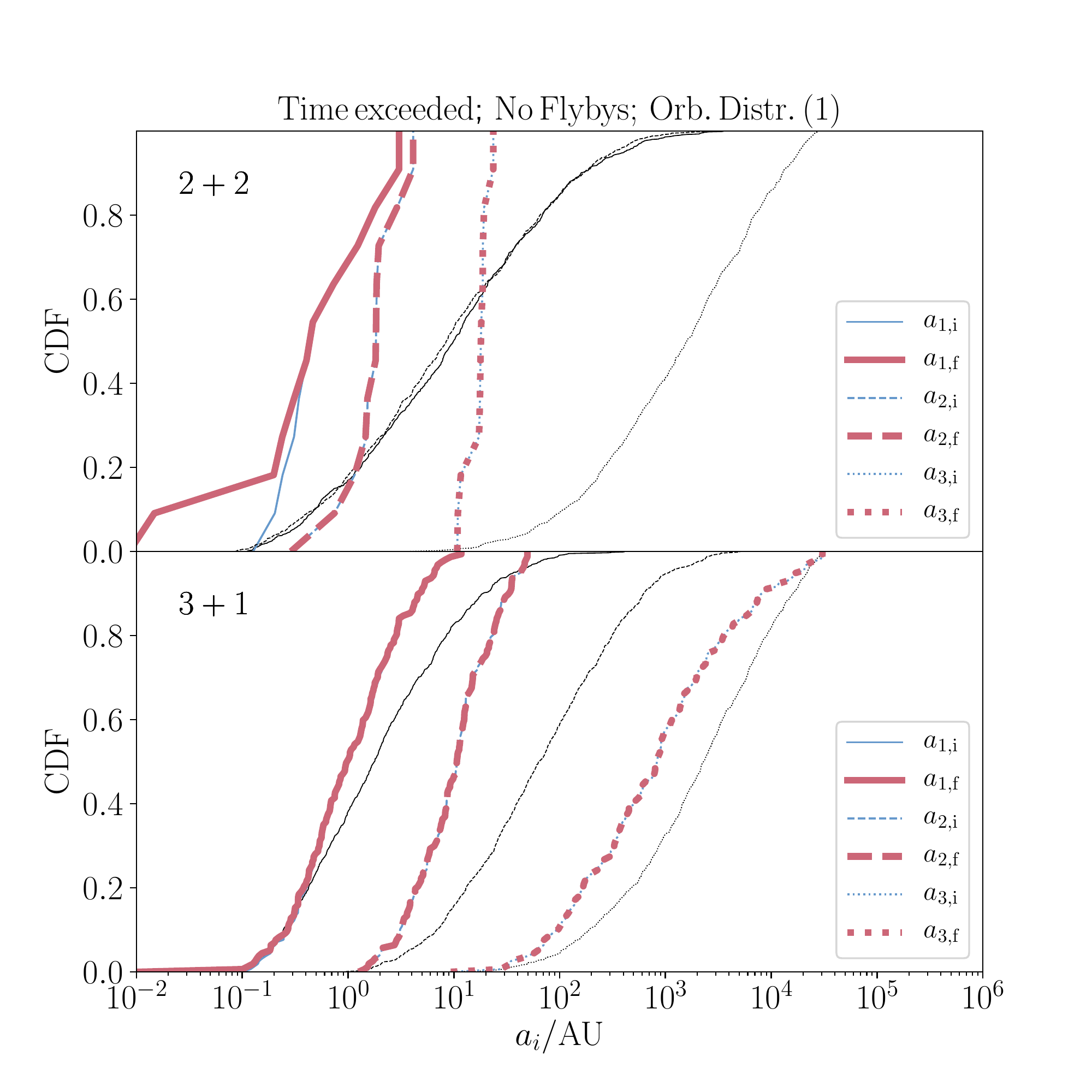}
\caption {Cumulative semimajor axis distributions of the migrating systems (orbit 1; top two panels) in the simulations with the first assumption about the orbital distributions, and no flybys included. Solid, dashed and dotted lines correspond to orbits 1, 2 and 3, respectively. The initial and final distributions are shown with the thin blue and thicker red lines, respectively. The initial distributions of {\it all} systems (i.e., not just the migrating systems) are shown with thin black lines. The bottom two panels correspond to the systems in which the CPU wall time was exceeded (see \S\,\ref{sect:discussion:exceed} for discussion). }
\label{fig:pop_syn_sma}
\end{figure}

\subsection{Other orbital properties}
\label{sect:pop_syn:orb}
Here, we discuss orbital properties other than pertaining to the innermost orbital periods. We focus on the simulations with the first assumption about the orbital distributions and with no flybys included. Generally, there are no major differences in these distributions between these assumptions.

\subsubsection{Semimajor axes}
\label{sect:pop_syn:orb:sma}
In \F\,\ref{fig:pop_syn_sma}, we show the cumulative semimajor axis distributions of the migrating systems (specifically, orbit 1; top two panels) in the simulations with the first assumption about the orbital distributions, and no flybys included. Solid, dashed and dotted lines correspond to orbits 1, 2 and 3, respectively. The initial and final distributions are shown with the thin blue and thicker red lines, respectively. The initial distributions of {\it all} systems (i.e., not just the migrating systems) are shown with thin black lines. The bottom two panels correspond to the systems in which the maximum CPU wall time was exceeded (see \S\,\ref{sect:discussion:exceed} for discussion).

Evidently, for the migrating systems, the final distribution of $a_1$ is peaked around a few $\times 10^{-2}$, corresponding to an orbital period of a few days. The initial $a_1$ of the migrating systems are somewhat smaller compared to all systems, which can be attributed to the fact that tighter orbits are more susceptible to tidal migration. For 2+2 systems, this translates into typically smaller $a_3$, to compensate for the smaller initial $a_1$. Note that, for the 2+2 systems, the initial and final distributions of $a_2$ are slightly distinct, with the final $a_2$ being slightly more compact. This can be attributed to tidal evolution in orbit 2. 

\begin{figure}
\center
\includegraphics[scale = 0.45, trim = 10mm -5mm 0mm 10mm]{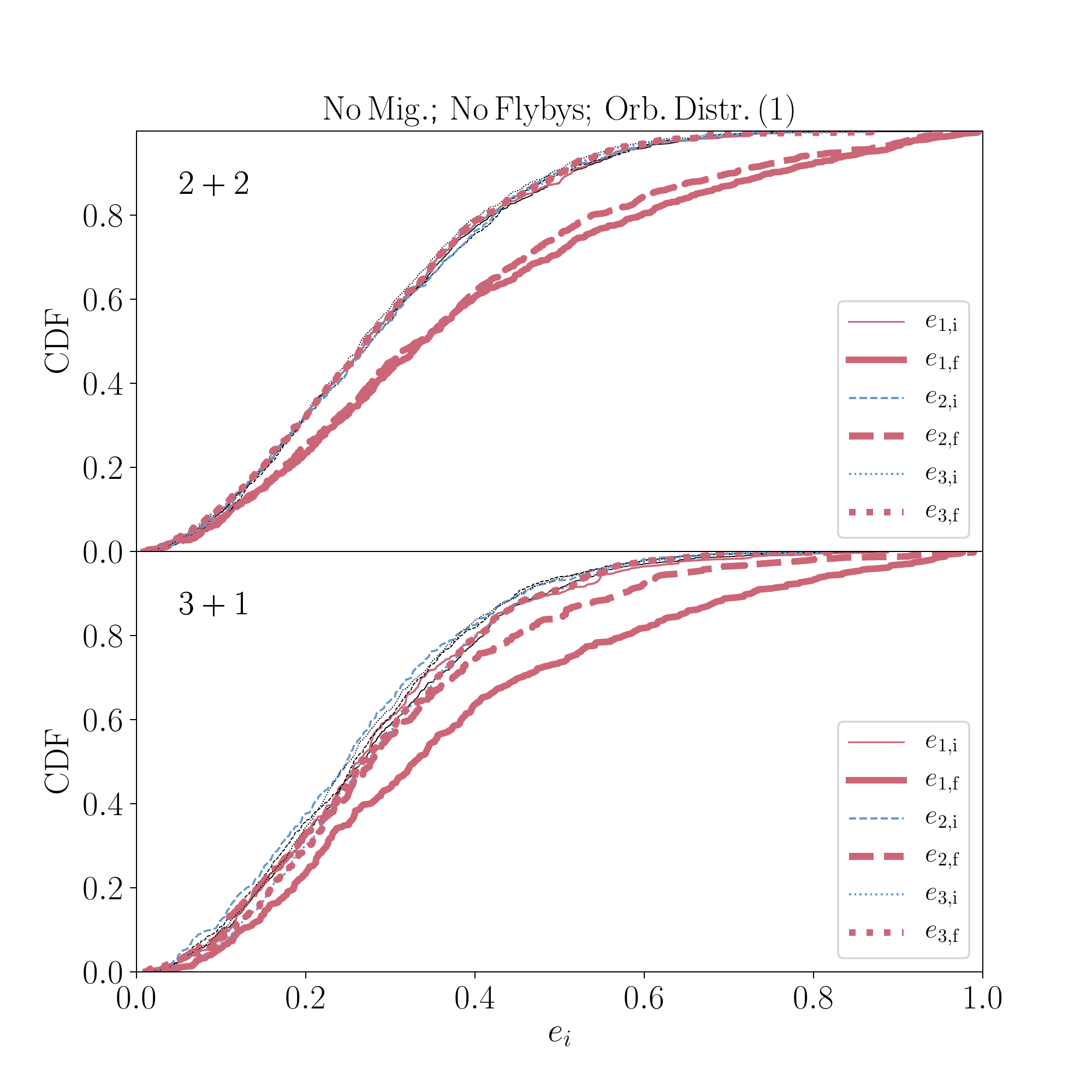}
\includegraphics[scale = 0.45, trim = 10mm 10mm 0mm 10mm]{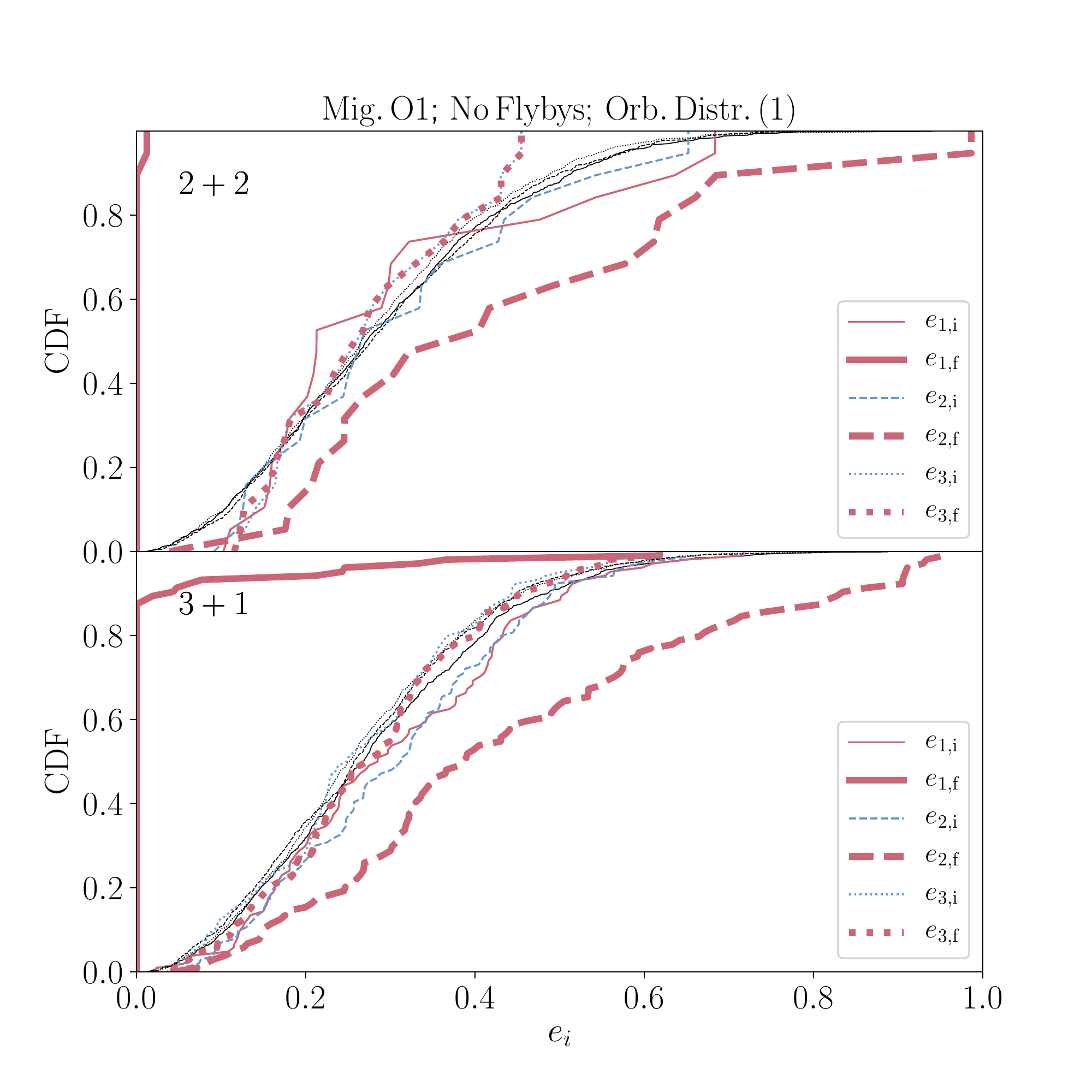}
\caption {Cumulative distributions of the eccentricities in the population synthesis simulations, similar to \F\,\ref{fig:pop_syn_sma}, here for the `no migration' (top two panels) and `orbit 1 migration' outcomes (bottom two panels). }
\label{fig:pop_syn_e}
\end{figure}

\subsubsection{Eccentricities}
\label{sect:pop_syn:orb:e}
In \F\,\ref{fig:pop_syn_e}, we show cumulative eccentricity distributions of the `no migration' (top two panels) and `orbit 1 migration' outcomes. The top two panels (`no migration') show that secular evolution can still significantly enhance the eccentricities of orbits 1 and 2 for both configurations. In the bottom two panels, the final distribution of $e_1$ is peaked around zero, with some exceptions (especially for the 3+1 configuration, for which the absolute number of migrating systems is larger). In those exceptions, tidal migration occurs, but the orbit is not yet completely circularized by 10 Gyr.

\begin{figure}
\center
\includegraphics[scale = 0.45, trim = 10mm 10mm 0mm 10mm]{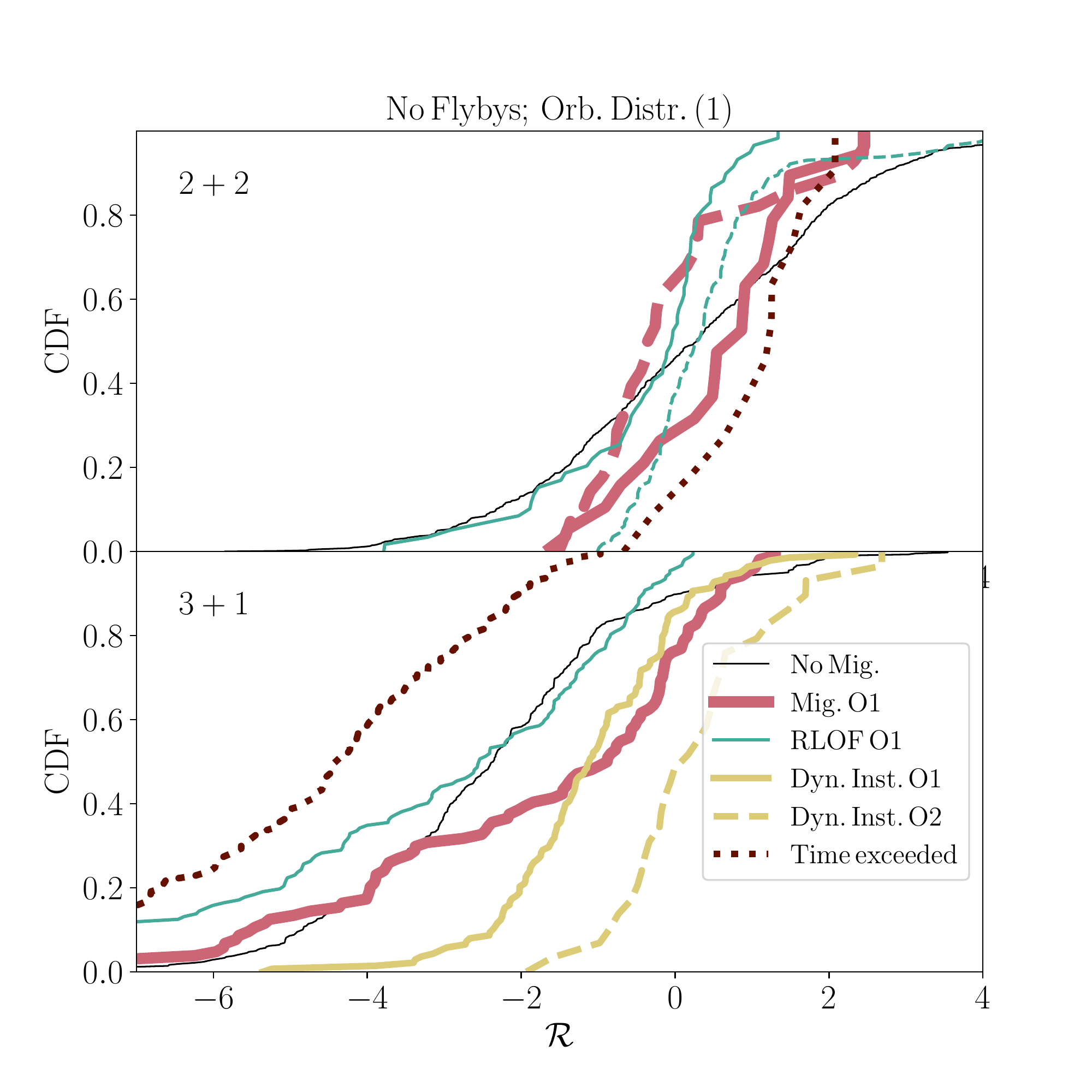}
\caption {Cumulative distributions of $\mathcal{R}_{2+2}$ (equation~\ref{eq:R_2p2}) and $\mathcal{R}_{3+1}$ (equation~\ref{eq:R_3p1}) in the top and bottom panels, respectively, for several outcomes in the simulations (indicated in the legend). The thin black lines show the initial distributions for {\it all} systems. }
\label{fig:pop_syn_R}
\end{figure}

\subsubsection{Ratios of LK time-scales}
\label{sect:pop_syn:orb:R}
As shown in previous studies \citep{2015MNRAS.449.4221H,2017MNRAS.470.1657H,2018MNRAS.474.3547G}, secularly chaotic behaviour and particularly high eccentricities are to be expected if the ratio of LK time-scales corresponding to a particular configuration is close to unity. In \F\,\ref{fig:pop_syn_R}, we show the cumulative distributions of $\mathcal{R}_{2+2}$ (equation~\ref{eq:R_2p2}) and $\mathcal{R}_{3+1}$ (equation~\ref{eq:R_3p1}) in the top and bottom panels, respectively, for several outcomes in the simulations (indicated in the legend). In particular, for the 2+2 systems, the migrating systems tend to have LK time-scale ratios that are more concentrated towards unity than the overall population. This applies similarly to the systems in which RLOF occurs, and is consistent with the expectation that eccentricity excitation peaks if the LK time-scales are comparable.

\subsubsection{Inclinations}

\begin{figure}
\center
\includegraphics[scale = 0.45, trim = 10mm -5mm 0mm 10mm]{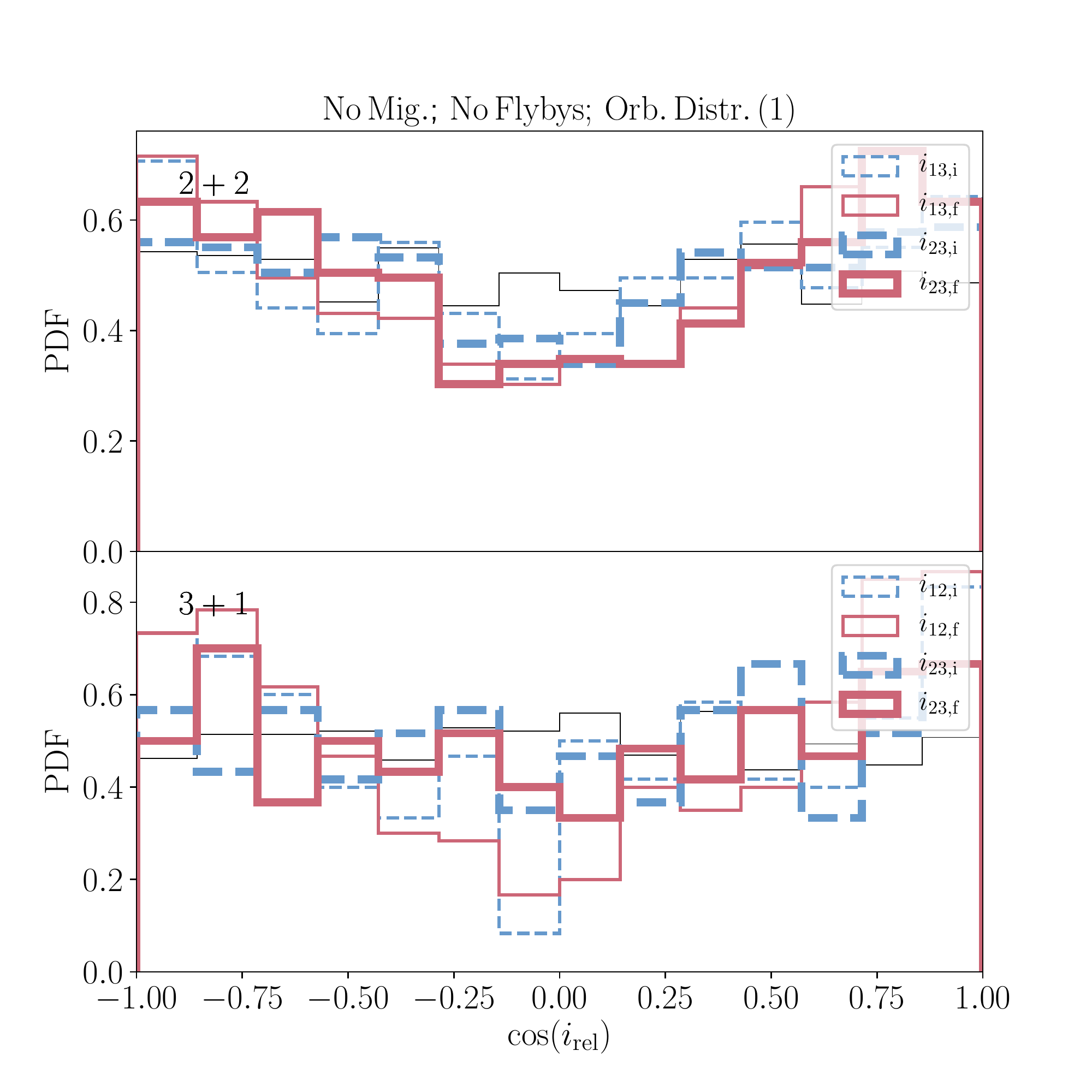}
\includegraphics[scale = 0.45, trim = 10mm 10mm 0mm 10mm]{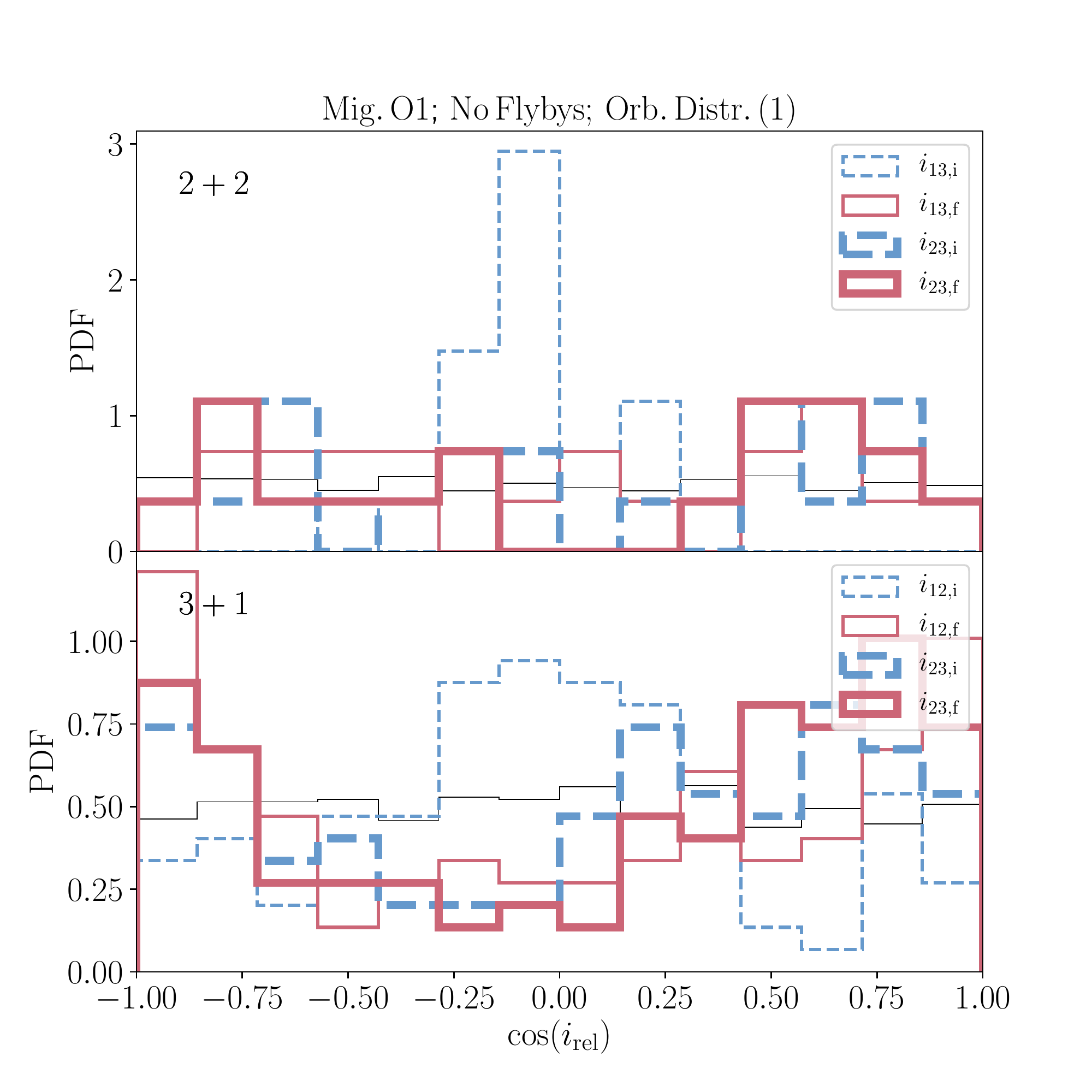}
\caption {Probability density distributions of the inclinations relative to the parent for the non-migrating (top two panels) and migrating (orbit 1; bottom two panels) systems in the population synthesis simulations. The initial (final) distributions are shown with solid (dashed) lines, respectively. The thin solid black lines show the initial distributions for all systems. }
\label{fig:pop_syn_incl}
\end{figure}

In \F\,\ref{fig:pop_syn_incl}, we show distributions of the inclinations relative to the parents of orbits 1 and 2 for the non-migrating (top two panels) and migrating (orbit 1; bottom two panels) systems. Note that for the 2+2 configuration, these inclinations are $i_{13}$ and $i_{23}$ for orbits 1 and 2, respectively; for the 3+1 configuration, they are $i_{12}$ and $i_{23}$. We recall that the initial orientations were assumed to be random, i.e., the initial distributions of the mutual inclinations were assumed to be flat in their cosine. 

The non-migrating systems show some paucity of final inner inclinations near $90^\circ$, and an enhancement near $\sim 50^\circ$ and $130^\circ$. This can be understood by noting that high inclinations trigger secular interactions, and in which case peaks in the distributions are expected around $50^\circ$ and $130^\circ$. In the case of migration of orbit 1, $i_{13,\mathrm{i}}$ ($i_{12,\mathrm{i}}$) tends to be concentrated around $90^\circ$ for the 2+2 (3+1) configurations, as expected based on three-body secular evolution. However, in particular for the 3+1 systems, $i_{12,\mathrm{i}}$ does not have to be close to $90^\circ$ --- values as small as a few tens of degrees are sufficient. This is consistent with \citet{2017MNRAS.470.1657H}, who showed that high eccentricities can be reached even for small initial mutual inclinations, provided that the ratio of LK time-scales is close to unity.

\subsection{Migration times}
\label{sect:pop_syn:time}
\begin{figure}
\center
\includegraphics[scale = 0.45, trim = 10mm 10mm 0mm 10mm]{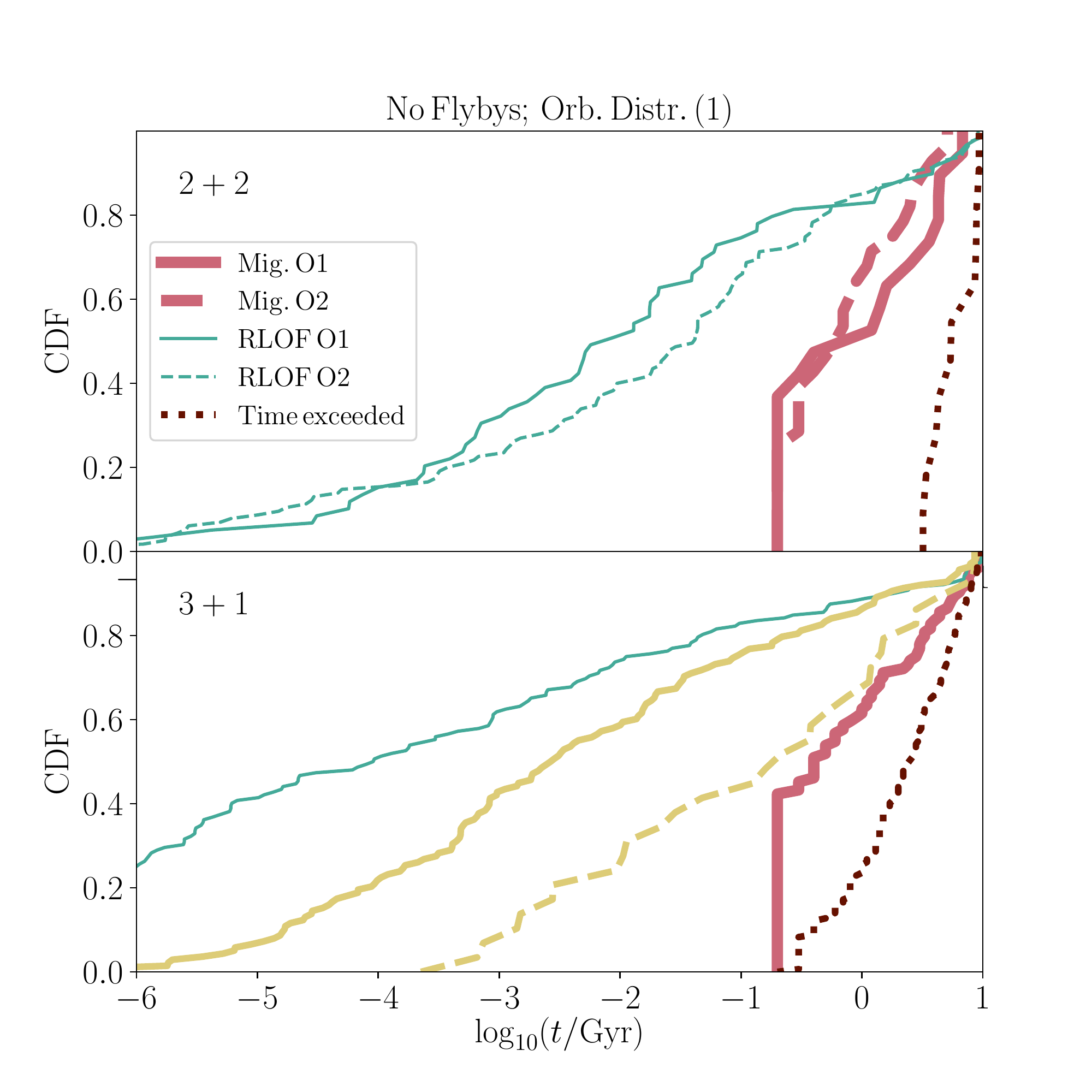}
\caption {Cumulative distributions of the migration times in the simulations for the 2+2 (top panel) and 3+1 (bottom panel) configurations. Also shown are cumulative distributions of the stopping times for various other outcomes. Refer to the legend for the meaning of the colours and line styles. }
\label{fig:stop_times}
\end{figure}

Lastly, we show in \F\,\ref{fig:stop_times} the cumulative distributions of the migration times for the migrating systems (orbit 1 or 2). We also show cumulative distributions of the stopping times for various other outcomes. The median migration time in our simulations is $\sim 1\,\mathrm{Gyr}$. Also shown are the times at which RLOF was triggered in orbit 1 or 2. RLOF is typically triggered early in the simulations, with a median time of $\sim 10\,\mathrm{Myr}$ and $\sim 0.1\,\mathrm{Myr}$ for the 2+2 and 3+1 configurations, respectively.

\section{Discussion}
\label{sect:discussion}

\subsection{Implications of the inner orbital period distributions}
\label{sect:discussion:per}
We found in \S\,\ref{sect:pop_syn:in} that our simulated inner orbital period distributions match the observations reasonably for 3+1 systems. However, the observations show the presence of a large population of 2+2 systems with inner periods around 10 d, which is not reproduced in the simulations. In our simulations, we made the simplifying assumption that no systems are formed at the MS with inner periods shorter than 10 d. This assumption is based on the argument that short-period systems would merge during their pre-MS evolution. However, the pre-MS evolution of short-period binaries in multiple systems is poorly understood. Evidently, if the initial inner orbital period distribution were closer to the observed distribution, then our simulations could be made to better match the observations. 

We do remark that for the 3+1 systems, the observed systems are reasonably described by secular and tidal evolution alone. This difference between the 2+2 and 3+1 systems could hint at different formation scenarios between 2+2 and 3+1 systems, i.e., sequential formation versus cascade (hierarchical) fragmentation (e.g., \citealt{2018AJ....155..160T}).

\subsection{Time-exceeded systems}
\label{sect:discussion:exceed}
As mentioned in Sections\,\ref{sect:meth:sc} and \ref{sect:pop_syn:frac}, in a number of systems, in particular for the 3+1 configuration, the integration proceeded very slowly. For practical reasons, the CPU wall time was limited to 24 hr. The fraction of terminated systems is less than 1 per cent for the 2+2 systems, but up to $\sim 16\%$ for the 3+1 systems. In the bottom two panels of \F\,\ref{fig:pop_syn_sma}, we show the distributions of the semimajor axes for the `time-exceeded' systems. For reference, we note that the top two panels of \F\,\ref{fig:pop_syn_sma} show the distributions for the migrating systems (orbit 1). The time-exceeded systems tend to have much tighter orbits compared to other systems, in particular the migrating systems. This can be understood by noting that the secular time-scales are shorter for more compact systems, implying that these systems are computationally demanding. Specifically, for the 2+2 configuration, $a_3\sim 20\au$ for the time-exceeded systems, which is significantly smaller than the typical $a_3$ for migrating systems ($a_3 \sim 3\times10^2\,\au$). For the 3+1 configuration, typically $a_2\sim 10\,\au$ for the time-exceeded systems, whereas $a_2\sim 10^2\,\au$ for the migrating systems. 

Other large differences between the migrating and time-exceeded systems are illustrated by the distributions of the LK time-scale ratios (\F\,\ref{fig:pop_syn_R}), and the migration/stopping times (\F\,\ref{fig:stop_times}). We conclude that the existence of the time-exceeded systems likely does not significantly affect our results of the orbital period distributions.

\subsection{RLOF}
\label{sect:discussion:RLOF}
We stopped our simulations at the onset of RLOF (see \S\,\ref{sect:meth:sc}). RLOF is expected to lead to mass transfer or common-envelope evolution, typically resulting in significant shrinkage of the orbit, or possibly even coalescence of two stars. The evolution can be complicated if the time-scales of mass transfer and secular evolution are comparable. Such evolution is beyond the scope of this work, but certainly merits further investigation. 

Nevertheless, we can remark that, if the result of RLOF is a tight orbit, then this would imply an enhancement of the inner orbital period distribution at short orbital periods ($<10\,\ud$). This could help to reduce tensions between the observed and simulated orbital period distributions (see \S\,\ref{sect:pop_syn:in}). On the other hand, coalescence of the stars would transform the system into a triple, and the observed inner orbital period distribution of quadruple stars would not apply in that case (however, it would affect the inner orbital period distribution of triple stars). 

Another related caveat is that the stellar radii may have been larger during the pre-MS phase (however, pre-MS phase evolution is still not fully understood, and some recent studies have shown that accretion during the pre-MS could affect the evolution compared to the standard picture of contraction along the Hayashi line, \citealt{1961PASJ...13..450H}, resulting in only modestly larger radii compared to the zero-age MS by a few tens of per cent, see, e.g., \citealt{2017A&A...599A..49K}). Larger radii during the pre-MS phase would trigger more RLOF in our simulations, in which zero-age MS radii were assumed. 

If RLOF were triggered during the pre-MS, then the stars would likely merge, and it would be more appropriate to identify the system as a triple system. Nevertheless, to investigate the potential effect of larger pre-MS radii, we carried out additional population synthesis simulations for a short duration of 1 Myr and with the primary star radius enlarged to $R_1=5 \,\rsun$ (taken to be constant during the 1 Myr integration). This enlarged radius is based on Fig. 2 of \citet{2017A&A...599A..49K}, and should give an upper limit to the effect of the larger primary star radius during the pre-MS (we do not consider a larger secondary star radius, since its mass and hence radius are smaller). The RLOF (star 1) fractions in these short pre-MS simulations are $\sim0.03$ ($\sim0.14$) for the 2+2 (3+1) systems. In simulations with a constant $R_1 = 1 \,\rsun$, the RLOF (star 1) fractions during the first 1 Myr are $\sim0.02$ ($\sim0.09$) for the 2+2 (3+1) systems. This implies that the larger primary radius increases the RLOF (star 1) fractions during the first Myr by a factor of $\sim1.5$, i.e., the pre-MS evolution does not affect the occurrence of RLOF by more than $\sim 50\%$.

\subsection{Breakdown of the averaging approximation}
\label{sect:discussion:sub}
The algorithm used in our integrations (see \S\,\ref{sect:meth:sec}) is based on an averaging of the Hamiltonian (and thus the equations of motion) over all three orbits. In reality, there exist short-term osculating eccentricity variations depending on the ratio of the outer orbital period to the secular time-scale (e.g., \citealt{2014MNRAS.439.1079A,2016MNRAS.458.3060L,2018MNRAS.476.4234F,2018MNRAS.481.4907G}). The averaging approximation can break down in various cases, for example when the time-scale for the angular momentum or eccentricity vector to change is comparable to some of the orbital periods \citep{2014ApJ...781...45A}. In addition, the inner orbit precession frequency can be close to, but still longer than the outer orbital period in some cases (such as in the Earth-Moon system), in which case evection terms may become important. In case of comparable inner orbit precession frequency and outer orbital period, the evection resonance can come into play (e.g., \citealt{2010A&A...515A..54F,2017MNRAS.466..276G}). Another example is the occurrence of mean-motion resonance in 2+2 systems \citep{2018MNRAS.475.5215B}. These effects are beyond the scope of this work.

\subsection{Galactic tides}
\label{sect:discussion:gal_tide}
We did not consider the effects of galactic tides in our simulations. The typical outermost orbit in our simulation has a semimajor axis of $\sim 10^3\,\au$, up to $\sim 10^4\,\au$ (see, e.g., the top two panels of \F\,\ref{fig:pop_syn_sma}), which is significantly below $10^5\,\au$, the separation at which galactic tides are expected to become important. We conclude that galactic tides are not important for the MS evolution of solar-type quadruple stars. However, we note that mass loss can drive orbital expansion in evolving quadruple star systems (e.g., \citealt{2018MNRAS.478..620H}). Therefore, galactic tides are potentially important in such systems.

\section{Conclusions}
\label{sect:conclusions}
We studied the formation of short-period orbits through tidal and secular evolution in hierarchical quadruple systems containing solar-type MS stars. We considered the 2+2 (two binaries orbiting each other's barycentre) and 3+1 (triple orbited by a fourth star) configurations (see \F\,\ref{fig:configurations}). In addition to secular gravitational and tidal evolution, we took into account the effects of encounters with passing stars. Our main conclusions are given below.

\medskip \noindent 1. In our simulations, the initial inner orbital periods were longer than 10 d. Due to secular and tidal evolution, the inner orbital periods shrank to $<10\,\ud$ in a few per cent of systems for the 2+2 configuration, and up to 14\% of systems for the 3+1 configuration. The higher migration efficiency for the 3+1 configuration can be attributed to typically tighter initial inner orbits, and typically stronger secular evolution for these systems. RLOF is triggered due to high eccentricity in up to $\sim15\%$ of systems, and occurs most frequently in the 3+1 systems. Dynamical instability of the system occurs most commonly for the 3+1 systems. For the latter, in most cases dynamical stability is triggered by increased eccentricity of the intermediate orbit (orbit 2) due to the secular torque of the outermost orbit (orbit 3). 

\medskip \noindent 2. Through examples, we have shown that, in the 2+2 configuration, tidal shrinkage of one orbit can trigger the second orbit to shrink as well, leading to two short-period orbits (`double migration'). For the 3+1 configuration, we have shown that the eccentricity of the intermediate orbit (orbit 2) can become enhanced in response to the shrinking of the innermost orbit, due to a reduction in the apsidal precession in orbit 2 imposed by orbit 1 \citep{2015MNRAS.449.4221H}. This can affect the subsequent evolution of the inner orbit (in particular, further shrinking the inner orbit), but also triggering dynamical instability of the system. This shows that dynamical instability in 3+1 systems can be triggered not only by mass loss in evolving systems (e.g., \citealt{2018MNRAS.478..620H}), but also due to tidal evolution during MS evolution.

\medskip \noindent 3. Our simulated inner orbital period distributions compare reasonably to the observations \citep{1997A&AS..124...75T,2018ApJS..235....6T} for the 3+1 systems. However, for 2+2 systems, the observed inner orbital period distribution shows a significant enhancement at $\sim 10\,\ud$, which is not reproduced by the simulations. This suggests that the inner orbital periods of 2+2 systems are not predominantly set by tidal and secular evolution, but by other processes, most likely occurring during the stellar formation and early evolution.

\medskip \noindent 4. The migrating systems in our simulations show a preference for similar LK time-scales in the appropriate orbit pairs, reflecting the fact that high eccentricities are induced due to coupled secular evolution in these cases.

\medskip \noindent 5. Our simulated inner orbital period distributions are not strongly dependent on whether or not flybys are taken into account. We also considered a set of simulations with different assumptions about the orbital distributions, which yielded no qualitatively significantly different results. Quantitatively, flybys can enhance (decrease) the migration fractions by a few tenths of per cent for the 2+2 (3+1) configuration. 

\section*{Acknowledgements}
I thank the anonymous referee for helpful comments. I gratefully acknowledge support from the Institute for Advanced Study, and The Peter Svennilson Membership.

\bibliographystyle{mnras}
\bibliography{literature}

\label{lastpage}
\end{document}